# Strengthening and toughening mechanisms in heterostructured laminates revealed by a phase field-enhanced crystal plasticity simulation

Yukai Xiong [a], Jianfeng Zhao [b,*], Jinling Liu [a], Jie Wang [c], Michael Zaiser [d,*], Xu Zhang [a,*]

[a] *Sichuan Province Key Laboratory of Advanced Structural Materials Mechanical Behavior and Service Safety, School of Mechanics and Aerospace Engineering, Southwest Jiaotong University, Chengdu 610031, China*

[b] *School of Materials Science and Engineering, Southwest Jiaotong University, Chengdu 610031, China*

[c] *Department of Engineering Mechanics, School of Aeronautics and Astronautics, Zhejiang University, Hangzhou, 310027, China*

[d] *WW8-Materials Simulation, Department of Materials Science, Friedrich-Alexander Universität Erlangen-Nürnberg, Dr.-Mack-Str. 77, Fürth 90762, Germany*

**Abstract**

Heterostructured (HS) materials exhibit excellent mechanical properties, combining high strength and significant ductility. Hetero-deformation-induced (HDI) hardening and strain de-localization are key to their strength-ductility synergy. However, existing models often fall short in addressing these aspects. In this work, a coupled framework integrating strain gradient crystal plasticity and phase field damage models is developed. The interface dominated HDI hardening in HS laminates is handled by introducing a heterogeneity coefficient into the back stress. The phase field model accounts for defect energy-driven damage and accurately represents the material's ductile damage behavior by accounting for effects of microstructure on crack initiation and propagation. Simulation results on HS laminates align well with experimental results and reflect the distribution of geometrically necessary dislocations and back stresses at interfaces between regions with dissimilar microstructure. Crack initiation and propagation are accurately described, providing valuable insights into fracture behavior. The model can predict how strength and ductility change upon variations of the HS laminate microstructure, thus providing an essential tool for microstructure optimization. This work enhances the understanding of deformation mechanisms in HS laminates and provides valuable insights for design and optimization of this class of materials.

| **Nomenclature** | |
| --- | --- |
| $\mathbf{F}$ | Deformation gradient |
| $\mathbf{F}_e$, $\mathbf{F}_p$ | Elastic deformation gradient, plastic deformation gradient |
| $\hat{\mathbf{L}}$ | Velocity gradient in the current configuration |
| $\hat{\mathbf{L}}_e$, $\hat{\mathbf{L}}_p$ | Elastic and plastic velocity gradient in the current configuration |
| $\mathbf{L}_p$ | Plastic velocity gradient in the intermediate configuration |
| $\mathbb{C}$, $\tilde{\mathbb{C}}$ | Elastic stiffness, elastic stiffness of the undamaged material |
| $\mathbf{E}_e$ | Elactic Green-Lagrange strain in the intermediate configuration |
| $\dot{\mathbf{E}}_e$ | Elactic Green-Lagrange strain rate |
| $\mathbf{P}$ | The first Piola-Kirchhoff stress |
| $\mathbf{S}$ | The second Piola-Kirchhoff stress |
| $\mathbf{S}_0$ | The second Piola-Kirchhoff stress in the undamaged state |
| $\mathbf{\Lambda}$ | Principle stress components |
| $\mathbf{Q}$ | Eigenvector matrix of the principle stress |
| $\mathbf{S}^+$, $\mathbf{S}^-$ | Tensile and compressive stress tensor |
| $\mathbb{P}^+$, $\mathbb{P}^-$ | Positive and negative components of the projection tensor |
| $\mathbf{u}$ | Displacement |
| $\mathbf{b}$ | Body force |
| $\boldsymbol{\xi}$ | Micro-force traction vector of the phase field |
| $f_i$ | Internal micro-force of the phase field |
| $\rho_0$ | Material density |
| $\mathbf{n}$ | Normal vector on the domain boundary $\partial \mathcal{B}_0$ |
| $r$, $\mathbf{q}$ | Heat source, heat flux |
| $e$, $\omega$ | Internal energy density, entropy density |
| $\underline{\rho}$ | A list of internal variables (dislocation densities) |
| $\theta$ | Temperature |
| $\psi$ | Helmholtz free energy density |
| $\psi_e$ | Elastic free energy density |
| $\psi_d$ | Defect free energy density |
| $\psi_f$ | Generalized fracture surface energy (damage energy density) |
| $\mathcal{H}$ | History field variable |
| $\varphi$, $\nabla \varphi$ | Damage variable, gradient of damage variable |
| $\bar{f}$ | Driving force of the phase field |
| $\Phi$ | Total dissipation rate |
| $\Phi_f$ | Damage/fracture dissipation rate |
| $\Phi_p$ | Plastic dissipation rate |
| $M_f$ | Mobility of the damage phase field |
| $\psi_{cr}$ | Critical defect energy density |
| $l_c$ | Characteristic width of the crack |
| $g_c$ | Critical energy release rate |
| $\dot{\gamma}$ | Shear slip rate |
| $\mathbf{m}^\alpha$ | Unit vector along the slip direction on $\alpha$ slip system |
| $\mathbf{n}^\alpha$ | Unit vector of slip-plane normal on $\alpha$ slip system |
| $\mathbf{t}^\alpha$ | Unit vector of tangential direction on $\alpha$ slip system |
| $\tau$ | Resolved shear stress |
| $\tau_{back}$, $\tau_{pass}$ | Back stress, passing stress |
| $\tau_{orowan}$, $\tau_{load}$ | Orowan strengthening stress, load transfer stress |
| $\tau_{eff}$, $\tau_P$ | Effective resolved shear stress, Peierls stress |
| $G$, $\upsilon$ | Shear modulus, Poisson's ratio |
| $b$ | Magnitude of the Burgers vector |
| $v_0$ | Dislocation glide velocity pre-factor |
| $Q_s$ | Activation energy |
| $k_B$ | Boltzmann constant |
| $p$, $q$ | Parameters control the glide resistance profile |
| $\rho_{SSD}$ | Statistically stored dislocation density |
| $\rho_{GND}$ | Geometrically necessary dislocation density |
| $L$ | Average particle spacing |
| $r_p$, $f_p$ | Particle radius, particle volume fraction |
| $\sigma_{int}$ | Intrinsic strengthening stress |
| $\vartheta$ | Effective aspect ratio of the particle |
| $\chi$ | Interaction strength coefficient |
| $\kappa$ | Taylor coefficient |
| $\rho_f$ | Forest dislocation density |
| $k_{mult}$ | Contribution coefficient of the forest dislocation |
| $d_{anni}$ | Critical annihilation distance |
| $\eta$ | Filter coefficient |
| $\lambda$ | Grid spacing |
| $D$ | Contribution coefficient of the back stress |
| $\delta_h$, $k_h$ | Heterogeneity coefficient, heterogeneity parameter |
| $r_h$ | Heterogeneous influence zone radius |
| $\mathbf{k}$, $k$ | Frequency vector and its magnitude in Fourier space |
| $\mathcal{F}[\cdot]$ | Fourier transform |
| $\mathcal{F}^{-1}[\cdot]$ | Inverse Fourier transform |

## 1. Introduction

Heterostructured (HS) materials are a class of materials composed of distinct regions with significantly differing mechanical or physical properties, often exhibiting enhanced functional and mechanical properties compared to homogeneous materials (Zhu and Wu, 2019, 2023). These materials encompass various configurations, including gradient structure (Fang et al., 2011; Lu, 2014; Wu et al., 2021; Wu et al., 2014), laminate structure (Beyerlein et al., 2014; Ma et al., 2016), bi-modal structure (Fan et al., 2006; Han et al., 2005; Nie et al., 2023; Zhang et al., 2019), harmonic structure (Li et al., 2021; Zhang et al., 2022), and metal matrix composites (Ma et al., 2023; Ramakrishnan, 1996; Samal et al., 2020).

During plastic deformation, HS materials exhibit non-uniform deformation at the interfaces between hard and soft domains, leading to the accumulation of geometrically necessary dislocations (GNDs) at these boundaries. The associated long-range internal stresses ('long-range' here refers to the scale of the grain microstructure as opposed to the much smaller scale of single dislocations and their spacing) gives rise to what is commonly referred to as back stress (Shukla et al., 2018; Yang et al., 2016). The exceptional mechanical properties of HS materials are primarily attributed to the back stress. The additional hardening observed in HS materials, arising from the build-up of back stress between the hard and soft domains, is denoted as hetero-deformation induced (HDI) hardening (Zhu and Wu, 2019). HDI hardening enhances the strength of the soft domains, leading to a higher overall yield strength than predicted by the rule-of-mixtures (ROM) (Fan et al., 2025). The back stress can be classified into intragranular and intergranular types (Chen et al., 2015). Both homogeneous and HS materials exhibit these forms of back stress. However, HS materials introduce microstructure heterogeneities above the grain scale. The resulting pronounced deformation heterogeneities may significantly enhance back stress over those present in homogeneous materials (HDI hardening). The back stress within HDI hardening is a dominant mechanism in enhancing strength and toughness of HS materials (Zhu and Wu, 2019).

Numerous experiments have confirmed that the heterogeneous configuration of HS materials can induce HDI stress and achieve an exceptional strength-ductility synergy (Huang et al., 2018; Liu et al., 2023; Nie et al., 2023; Shukla et al., 2018; Wang et al., 2023; Wu et al., 2015; Wu et al., 2025). The heterogeneity in HS materials spans multiple length scales: at the microscale, it is manifested through variations in dislocation structures, twinning, and second-phase particles, whereas at the mesoscale, it is reflected in differences in grain morphology and the local mechanical properties of constituent domains. For example, dual-structure titanium composites demonstrated high strength while maintaining significant ductility. Reducing the size of heterogeneous phases and thus increasing the number of hetero-interfaces (HIs) enhanced their fracture resistance (Liu et al., 2023). In HS materials composed of soft coarse grained (CG) domains and hard ultrafine-grained domeins, materials with less pronounced heterogeneity due to smaller grain sizes in the soft domain and thinner soft domain regions exhibited reduced fracture resistance (Nie et al., 2023). Microcracks were found to typically initiate in high-stress areas of the hard domains or at weak interfaces, and were blunted and arrested by the soft domains (Fan et al., 2022).

Interfaces between hard and soft regions play a key role in controlling strength and ductility of HS materials. While the proportion of heterogeneous interface (HI) influences HDI hardening, an excessive increase at HI can significantly reduce ductility (Wu et al., 2015; Zhu et al., 2021). Current design principles for optimizing heterogeneous microstructures primarily focus on the dimensions of interface-affected zones (IAZ) and use considerations of toughening based on linear elastic fracture mechanics (Fan et al., 2022). Experiments suggest that optimal strength-ductility matching is achieved when the width of CG layers in heterogeneous laminates is twice the IAZ width (Fu et al., 2021) or when the width of CG regions approximates the width of the plastic zone around crack tips (Liu et al., 2020; Ma et al., 2021). However, the physical principles underlying such phenomenological relations between microstructure geometry and mechanical properties require further clarification.

To investigate the mehanisms governing strength and ductility of HS materials at the grain scale, the crystal plasticity finite element method (CPFEM) is widely employed to simulate their elastic-plastic deformation behavior (Ardeljan et al., 2014; Guo et al., 2024; Zhang et al., 2023; Zhang et al., 2021). In CP constitutive models, spatial couplings associated with non-uniform deformation are described in terms of densities of GNDs and their gradients, whose evolution in turn derives from the spatially non-uniform deformation rates. Terms containing GND densites and density gradients thus connect the mechanical responses of material points to those of their neighbors (Gao and Huang, 2003; Ma et al., 2006). Zhang et al. (2023) explored GND hardening and back stress strengthening in gradient materials using a nonlocal constitutive model incorporating dislocation fluxes. Guo et al. (2024) investigated the strengthening mechanisms in gradient nano-twinned copper using a strain gradient model based on dislocation-twin interactions. However, the effectiveness of strain gradient models in capturing nonlocal behavior on larger scales, particularly the HDI effect, has yet to be fully established.

This study aims to describe the plastic behavior of HS laminates using a strain gradient CP model. However, Cheong's research on FCC polycrystals with strain gradient models has highlighted significant mesh sensitivity issues (Cheong et al., 2005). This sensitivity is especially pronounced in polycrystal models with small grain sizes, where finer meshes lead to higher stress values, and this phenomenon does not converge with mesh refinement. This indicates the appearance of spurious strain heterogeneities on the scale of individual mesh elements, leading to unphysical accumulation of GNDs. To address this issue, numerical regularization approaches are needed to suppress short-wavelength discretization artefacts. For example, Wulfinghoff and Böhlke (2015) introduced a diffusion term into the equations describing dislocation density transport. Here, we will explore alternative approaches to obtain a mesh insensitive numerical formulation.

Most simulation studies on HS materials have focused on their hardening behavior, leaving the specifics of crack propagation in heterostructures relatively unexplored. While RVE simulations provide insight into microstructural plasticity and relative ductility, they are limited in capturing macroscopic crack propagation and fracture due to the small model size. Within this framework, strength is characterized by the yield and flow stresses, while ductility is interpreted as the plastic strain accumulated before the onset of significant softening or microstructure-level deformation localization. The simulations capture the nucleation and early propagation of microcracks, which are closely related to the initiation of localization observed experimentally, thereby providing a physically meaningful microscale measure of ductility. The phase-field (PF) method is widely employed to investigate damage and fracture behavior in materials and attempts have been made to adapt it to ductile behavior and inhomogeneous microstructures (Salvini et al., 2024; Shanthraj et al., 2016; Shanthraj et al., 2017; Wu et al., 2020). This method substitutes discontinuous crack surfaces with continuous damage variables, effectively mitigating crack tip singularities. However, the majority of work on PF fracture has focused on modesl of brittle behavior. Such models cannot capture the interplay between microstructure-level plasticity and crack propagation, for which an adequate coupling between PF damage and plasticity is essential. There exist PF models for ductile failure which may act as useful starting points, see e.g. Ambati et al. (2015); (2016) and Shanthraj et al. (2016); (2017)

but further work is needed to adopt these models to the highly heterogeneous deformation modes and defect distributions typical of HS materials.

The remainder of this paper is organized as follows: Section 2 introduces the constitutive model coupling strain gradient CP and PF, including the discussion of a novel approach to adopt PF models of damage accumulation to HS materials. Section 3 details the parameter configuration for the coupled model applied to HS materials and the validation of the constitutive model. Section 4 elucidates the deformation mechanisms of HS laminates on the microstructure level. Section 5 investigates the correlations between microstructure and macroscopic mechanical properties. Finally, Section 6 summarizes the key findings. Additional supporting information is provided in the Appendix.

## 2. Theory

A schematic diagram of the CP-PF coupling used in this work is shown in Fig. 1. The CP module describes the evolution of dislocations and transmits the system's free energy $\psi$, calculated based on the stress and strain states, to the PF module. The PF module solves the evolution of the damage variable $\varphi$ and transmits the damage state back to the CP module.

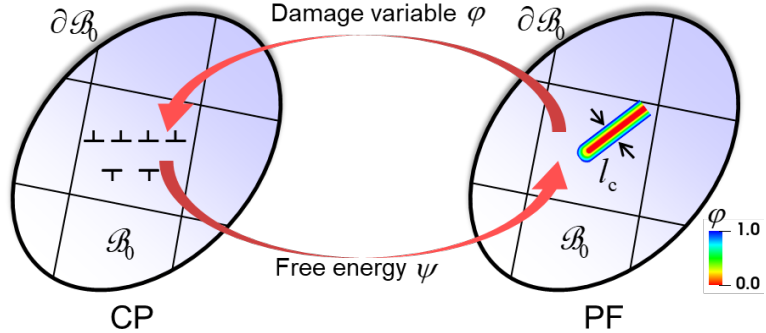

**Fig. 1.** The CP-PF coupled model transfers variables as follows: CP describes the evolution of dislocations and transfers the free energy to PF, which then calculates the evolution of the damage order parameter describing the diffuse crack surface of width $l_c$.

### 2.1 Kinematics

We consider a stress-free, undeformed reference configuration $\mathcal{B}_0 \subset \mathbb{R}^3$ with boundary $\partial \mathcal{B}_0$, the deformation resulting from an applied loading defines a field $\chi(\mathbf{x})$ mapping the material point $\mathbf{x}$ in the reference configuration $\mathcal{B}_0$ to point $y$ in the current, deformed configuration $\mathcal{B}$. The total deformation gradient $\mathbf{F}$, given by $\mathbf{F} = \partial \chi / \partial \mathbf{x}$, is multiplicatively decomposed into elastic deformation gradient $\mathbf{F}_e$ and plastic deformation gradient $\mathbf{F}_p$

$$\mathbf{F} = \mathbf{F}_e \cdot \mathbf{F}_p. \tag{1}$$

Here, $\mathbf{F}_p$ is a lattice-preserving, inelastic deformation gradient that maps from the reference configuration to a stress-free, plastically deformed intermediate configuration, and the elastic deformation gradient $\mathbf{F}_e$ maps from the intermediate to the current configuration.

The velocity gradient $\hat{\mathbf{L}}$ in the current configuration is obtained by taking the time derivative of this equation and decomposing it into an elastic part $\hat{\mathbf{L}}_e$ and plastic part $\hat{\mathbf{L}}_p$,

$$\hat{\mathbf{L}} = \dot{\mathbf{F}} \cdot \mathbf{F}^{-1} = \dot{\mathbf{F}}_e \cdot \mathbf{F}_e^{-1} + \mathbf{F}_e \cdot \left( \dot{\mathbf{F}}_p \cdot \mathbf{F}_p^{-1} \right) \cdot \mathbf{F}_e^{-1}, \tag{2}$$

$$\hat{\mathbf{L}}_e = \dot{\mathbf{F}}_e \cdot \mathbf{F}_e^{-1}, \tag{3}$$

$$\hat{\mathbf{L}}_p = \mathbf{F}_e \cdot \left( \dot{\mathbf{F}}_p \cdot \mathbf{F}_p^{-1} \right) \cdot \mathbf{F}_e^{-1}, \tag{4}$$

$$\mathbf{L}_p = \dot{\mathbf{F}}_p \cdot \mathbf{F}_p^{-1}. \tag{5}$$

Here $\mathbf{L}_p$ represents the plastic velocity gradient in the intermediate configuration.

The elastic response of the material is described by the generalized Hooke law

$$\mathbf{S} = \mathbb{C} : \mathbf{E}_e, \tag{6}$$

where $\mathbb{C}$ is the elastic stiffness tensor, $\mathbf{S}$ is the the second Piola-Kirchhoff stress in the intermediate configuration. The elastic Green-Lagrange strain $\mathbf{E}_e$ is given by $\mathbf{E}_e = \left( \mathbf{F}_e^T \cdot \mathbf{F}_e - \mathbf{I} \right)/2$ with the identity matrix $\mathbf{I}$ in the intermediate configuration.

**2.2 Balance laws and entropy inequality**

2.2.1 Momentum conservation

From conservation of linear and angular momentum, the following force equilibrium equation can be derived:

$$\rho_0 \ddot{\mathbf{u}} = \nabla \cdot \mathbf{P} + \mathbf{b}, \quad \mathbf{F} \cdot \mathbf{P}^T = \mathbf{F}^T \cdot \mathbf{P}, \tag{7}$$

where $\rho_0$ is the initial material density, $\mathbf{u}$ is the displacement, and $\mathbf{b}$ is the body force per reference volume.

The microscopic state of the material is described in terms of a damage parameter $\varphi$, which we define in such a manner that a value $\varphi = 1$ corresponds to an undamaged state and $\varphi = 0$ represents complete damage. The evolution of the damage parameter $\varphi$ can be considered in terms of a balance of micro-forces (Gurtin, 1996; McAuliffe and Waisman, 2015; Svolos et al., 2025; Zeng et al., 2022). Therefore, the corresponding local equilibrium equation

is given by

$$\nabla \cdot \boldsymbol{\xi} + f_i = 0, \tag{8}$$

where $\boldsymbol{\xi}$ is the micro-force traction vector, $f_i$ is the internal micro-force.

2.2.2 The first law of thermodynamics

According to the first law of thermodynamics, the energy conservation equation can be expressed as

$$\frac{d}{dt}\int_{\mathcal{B}_0}\left(\frac{1}{2}\rho_0\|\dot{\mathbf{u}}\|^2 + \rho_0 e\right)dv = \int_{\partial\mathcal{B}_0}(\mathbf{P}\cdot\mathbf{n})\cdot\dot{\mathbf{u}}\ ds + \int_{\mathcal{B}_0}\mathbf{b}\cdot\dot{\mathbf{u}}\ dv \\ + \int_{\partial\mathcal{B}_0}(\boldsymbol{\xi}\cdot\mathbf{n})\dot{\varphi}\ ds + \int_{\mathcal{B}_0}\rho_0 r\ dv - \int_{\partial\mathcal{B}_0}\mathbf{n}\cdot\mathbf{q}\ ds, \tag{9}$$

where $e$ represents the internal energy per unit mass, $\mathbf{n}$ is the normal vector on the domain boundary $\partial\mathcal{B}_0$, $r$ is the heat source, and $\mathbf{q}$ is the heat flux. The rate of change of internal energy per unit volume $\rho_0 \dot{e}$ can be expressed as

$$\rho_0 \dot{e} = \mathbf{P}:\dot{\mathbf{F}} + \boldsymbol{\xi}\cdot\nabla\dot{\varphi} - f_i\dot{\varphi} + \rho_0 r - \nabla\cdot\mathbf{q}. \tag{10}$$

2.2.3 The second law of thermodynamics

According to the second law of thermodynamics, the Clausius-Duhem inequality can be described as

$$\rho_0 \dot{\omega} \geq -\nabla\cdot\frac{\mathbf{q}}{\theta} + \rho_0\frac{r}{\theta}, \tag{11}$$

where $\omega$ represents the entropy density per unit mass and $\theta$ is temperature. Defining the Helmholtz free energy density $\psi = e - \theta\omega$, the corresponding rate can be expressed as $\dot{\psi} = \dot{e} - \dot{\theta}\omega - \theta\dot{\omega}$. Assuming the system is isothermal, quasi-static, and without external heat source, and substituting Eq. (10) into Eq. (11), we obtain

$$\mathbf{P}:\dot{\mathbf{F}} + \boldsymbol{\xi}\cdot\nabla\dot{\varphi} - f_i\dot{\varphi} - \rho_0\dot{\psi} \geq 0. \tag{12}$$

The free energy densitry $\psi$ is related to the elastic Green-Lagrange strain $\mathbf{E}_e$, the damage variable $\varphi$, the spatial gradient of the damage variable $\nabla\varphi$, and a set of internal variables $\underline{\rho}$ describing the defect state (in the present context: dislocation microstructure) of the material. The corresponding equation can be expressed as

$$\psi = \psi\left(\mathbf{E}_e,\ \varphi,\ \nabla\varphi,\ \underline{\rho}\right). \tag{13}$$

The time derivative of the free energy density can be derived as

$$\dot{\psi} = \dot{\psi}\left(\mathbf{E}_e,\ \varphi,\ \nabla\varphi,\ \underline{\rho}\right) = \frac{\partial\psi}{\partial\mathbf{E}_e}:\dot{\mathbf{E}}_e + \frac{\partial\psi}{\partial\varphi}\dot{\varphi} + \frac{\partial\psi}{\partial\nabla\varphi}\cdot\nabla\dot{\varphi} + \frac{\partial\psi}{\partial\underline{\rho}}\cdot\dot{\underline{\rho}}. \tag{14}$$

Based on the power conjugation relationship, $\mathbf{P}:\dot{\mathbf{F}} = \mathbf{S}:\dot{\mathbf{E}}_e + \mathbf{S}:\mathbf{L}_p$, the dissipation inequality for the system can be further written as

$$\Phi = \mathbf{S}:\dot{\mathbf{E}}_e + \mathbf{S}:\mathbf{L}_p + \boldsymbol{\xi}\cdot\nabla\dot{\varphi} - f_i\dot{\varphi} - \rho_0\dot{\psi} \geq 0. \tag{15}$$

Substituting Eq. (14) into Eq. (15), the dissipation inequality is given by

$$\Phi = \left(\mathbf{S} - \rho_0\frac{\partial\psi}{\partial\mathbf{E}_e}\right):\dot{\mathbf{E}}_e + \mathbf{S}:\mathbf{L}_p + \left(-f_i - \rho_0\frac{\partial\psi}{\partial\varphi}\right)\dot{\varphi} + \left(\boldsymbol{\xi} - \rho_0\frac{\partial\psi}{\partial\nabla\varphi}\right)\cdot\nabla\dot{\varphi} - \rho_0\frac{\partial\psi}{\partial\underline{\rho}}\cdot\dot{\underline{\rho}} \geq 0. \tag{16}$$

2.2.4 Constitutive relations

In the above equation, the elastic strain rate $\dot{\mathbf{E}}_e$, the damage evolution rate $\dot{\varphi}$, and the rate of the damage gradient $\nabla\dot{\varphi}$ can take arbitrary values. Therefore, the following relationship can be derived

$$\mathbf{S} = \rho_0\frac{\partial\psi}{\partial\mathbf{E}_e}, \tag{17}$$

$$\boldsymbol{\xi} = \rho_0\frac{\partial\psi}{\partial\nabla\varphi}. \tag{18}$$

In accordance with Eq. (16), the fracture dissipation $\Phi_f$ and plastic dissipation $\Phi_p$ are required to be non-negative

$$\Phi_f = \left(-f_i - \rho_0\frac{\partial\psi}{\partial\varphi}\right)\dot{\varphi} \geq 0, \tag{19}$$

$$\Phi_p = \mathbf{S}:\mathbf{L}_p - \rho_0\frac{\partial\psi}{\partial\underline{\rho}}\cdot\dot{\underline{\rho}} \geq 0. \tag{20}$$

Combining Eqs. (18) and (8), we can obtain

$$f_i = -\nabla\cdot\left(\rho_0\frac{\partial\psi}{\partial\nabla\varphi}\right). \tag{21}$$

Based on Eqs. (19) and (21), we define the thermodynamic driving force of the damage variable $\varphi$ as

$$\overline{f} = -f_i - \rho_0\frac{\partial\psi}{\partial\varphi} = \nabla\cdot\left(\rho_0\frac{\partial\psi}{\partial\nabla\varphi}\right) - \rho_0\frac{\partial\psi}{\partial\varphi}. \tag{22}$$

The evolution of the damage variable is governed by a Ginzburg-Landau-type equation

$$\dot{\varphi} = M_f\overline{f}, \tag{23}$$

where $M_f$ is the positive mobility.

Substituting the thermodynamic driving force $\overline{f}$ of the damage variable from Eq. (22) into Eq. (23), we obtain

$$\frac{\dot{\varphi}}{M_{\mathrm{f}}} = -\left(\rho_0 \frac{\partial \psi}{\partial \varphi} - \rho_0 \nabla \cdot \frac{\partial \psi}{\partial \nabla \varphi}\right). \tag{24}$$

Combining Eqs. (19), (22), and (23), we obtain that the damage dissipation $\Phi_{\mathrm{f}}$ satisfies the dissipation inequality

$$\Phi_{\mathrm{f}} = \bar{f}\dot{\varphi} = \frac{\dot{\varphi}^2}{M_{\mathrm{f}}} \geq 0. \tag{25}$$

The plastic energy is entirely dissipated as heat, except for the portion driving the evolution of a set of internal variables (dislocation densities) $\underline{\rho}$. Therefore, the plastic dissipation $\Phi_{\mathrm{p}}$ fulfills the dissipation inequality

$$\Phi_{\mathrm{p}} = \mathbf{S}:\mathbf{L}_{\mathrm{p}} - \rho_0 \frac{\partial \psi}{\partial \underline{\rho}} \cdot \underline{\dot{\rho}} > 0. \tag{26}$$

**2.3 Phase field model**

In the constitutive model for damage evolution, the total free energy is decomposed into elastic $\psi_{\mathrm{e}}$, defect $\psi_{\mathrm{d}}$ and damage/fracture $\psi_{\mathrm{f}}$ contributions

$$\rho_0 \psi = \rho_0 \psi_{\mathrm{e}} + \rho_0 \psi_{\mathrm{d}} + \rho_0 \psi_{\mathrm{f}}. \tag{27}$$

Given that fracture in most metals is primarily driven by tensile stress, we decompose the elastic strain energy into tensile $\psi_{\mathrm{e}}^{+}$ and compressive parts $\psi_{\mathrm{e}}^{-}$, and apply the degradation function $g(\varphi)$ solely to the tensile contribution (Miehe et al., 2010; Zhang et al., 2020)

$$\rho_0 \psi_{\mathrm{e}} = g(\varphi)\psi_{\mathrm{e}}^{+} + \psi_{\mathrm{e}}^{-}. \tag{28}$$

We adopt the spectral decomposition of the undamaged second Piola-Kirchhoff stress $\mathbf{S}_0$ to determine the values of $\psi_{\mathrm{e}}^{+}$ and $\psi_{\mathrm{e}}^{-}$

$$\mathbf{S}_0 = \tilde{\mathbb{C}}:\mathbf{E}_{\mathrm{e}} = \mathbf{Q}\mathbf{\Lambda}\mathbf{Q}^{\mathrm{T}}, \tag{29}$$

where $\tilde{\mathbb{C}}$ is the stiffness of the undamaged material, $\mathbf{\Lambda} = \mathrm{diag}(\lambda_1, \lambda_2, \lambda_3)$ is a diagonal matrix composed of the three eigenvalues of the stress tensor, and $\mathbf{Q}$ is the eigenvector of the stress tensor. Therefore, the principal stress tensors for tension and compression can be expressed as

$$\mathbf{\Lambda}^{+} = \mathbf{I}^{+}\mathbf{\Lambda}\mathbf{I}^{+}, \ \mathbf{\Lambda}^{-} = \mathbf{I}^{-}\mathbf{\Lambda}\mathbf{I}^{-}, \tag{30}$$

where $\mathbf{I}^{+}$ satisfies $I_{ii}^{+} = 0$ when $\lambda_i < 0$, and $\mathbf{I}^{-}$ satisfies $I_{ii}^{-} = 0$ when $\lambda_i > 0$.

Thus, the tensile and compressive stress tensors can be expressed as

$$\mathbf{S}^{+} = \mathbf{Q}\mathbf{\Lambda}^{+}\mathbf{Q}^{\mathrm{T}}, \ \mathbf{S}^{-} = \mathbf{Q}\mathbf{\Lambda}^{-}\mathbf{Q}^{\mathrm{T}}, \tag{31}$$

and consequently, the free energy can be derived as

$$\psi_e^+ = \frac{1}{2}\mathbf{S}^+ : \mathbf{E}_e, \quad \psi_e^- = \frac{1}{2}\mathbf{S}^- : \mathbf{E}_e. \tag{32}$$

We can define projection tensors to decompose the stress into tensile and compressive stresses (Miehe, 1998)

$$\mathbf{S}^+ = \mathbb{P}^+ : \mathbf{S}_0, \quad \mathbf{S}^- = \mathbb{P}^- : \mathbf{S}_0, \tag{33}$$

where the projection tensors $\mathbb{P}^+$ and $\mathbb{P}^-$ are calculated as

$$\mathbb{P}^+ = \frac{\partial \mathbf{S}^+}{\partial \mathbf{S}_0}, \quad \mathbb{P}^- = \mathbb{I} - \mathbb{P}^+, \tag{34}$$

with the fourth-order identity tensor $\mathbb{I}$.

Combining Eqs. (6), (17) and (27), the elastic stiffness can be expressed as

$$\mathbb{C} = g(\varphi)\mathbb{P}^+ : \tilde{\mathbb{C}} + \mathbb{P}^- : \tilde{\mathbb{C}}. \tag{35}$$

The evolution of damage in plastically deforming materials is affected by the presence of plasticity induced defects. Ambati et al. (Ambati et al., 2015; Ambati et al., 2016) formulated a degradation function $g(\varphi)$ for ductile fracture where the internal variable controlling damage evolution is taken to be the equivalent plastic strain. However, this approach may not be fully appropriate for polycrystals with pronounced microstructural heterogeneity, where the accumulation of defects is strongly heterogeneous whereas plastic strain after prolonged deformation exhibits only weak heterogeneity. Moreover, because of the piling-up of dislocations, the locations of maximum plastic strain do not always coincide with the locations of maximum defect concentration. To capture the influence of plasticity-induced defects on damage, we therefore use the accumulated defect energy $\psi_d$ as the controlling parameter for damage evolution

$$g(\varphi) = \varphi^{2p} \quad \text{with} \quad p = \rho_0 \psi_d / \psi_{cr}. \tag{36}$$

The normalization constant $\psi_{cr}$ defines a critical value of the accumulated defect energy above which damage becomes pronounced. In a line energy approximation with dislocation line energy $Gb^2$, the defect energy of the dislocations can be expressed as

$$\rho_0 \psi_d = Gb^2 \sum_\beta \left( \rho_{SSD}^\beta + \rho_{GND}^\beta \right), \tag{37}$$

where $G$ represents the shear modulus, $b$ is the magnitude of the Burgers vector, and $\rho_{SSD}^\beta$ and $\rho_{GND}^\beta$ are the statistically stored and geometrically necessary dislocation densities on slip system $\beta$. The equations of evolution of these quantities are discussed in Section 2.4.

To guarantee crack irreversibility, we adopt a history field variable $\mathcal{H}(t)$ formulated as

$$\mathcal{H}(t) = \max_{\tau \in [0,\,t]} \{\psi_e^+(\tau)\}. \tag{38}$$

The damage contribution to the free energy is given by the PF approximation of the energy of a discrete crack surface, which includes both a homogeneous and gradient-dependent components (Shanthraj et al., 2016)

$$\rho_0 \psi_f = \frac{g_c}{l_c}(1-\varphi) + \frac{1}{2}g_c l_c |\nabla \varphi|^2, \tag{39}$$

where $l_c$ is the characteristic width scale that regularizes the damage field and controls the width of the continuous transition zone between undamaged and fully damaged material. $g_c$ is the critical energy release rate. Interface energy scaling of the damage PF model is described in detail in Appendix A.

By combining Eqs. (24) and (27), the evolution of the damage variable $\varphi$ can be written as

$$\dot{\varphi} = -M_f \left[ 2p\varphi^{2p-1}\mathcal{H} - \frac{g_c}{l_c} - \mathrm{div}(g_c l_c \nabla \varphi) \right]. \tag{40}$$

To ensure the stability of the numerical solution, the minimum value of the damage variable is constrained to 0.001.

**2.4 Crystal plasticity model**

The plastic velocity gradient in the intermediate configuration $\mathbf{L}_p$ is calculated from the slip rates $\dot{\gamma}^\alpha$ on a set of slip systems $\alpha$

$$\mathbf{L}_p = \dot{\mathbf{F}}_p \cdot \mathbf{F}_p^{-1} = \sum_\alpha \dot{\gamma}^\alpha \left( \mathbf{m}^\alpha \otimes \mathbf{n}^\alpha \right), \tag{41}$$

where $\mathbf{m}^\alpha$ and $\mathbf{n}^\alpha$ are unit vectors along the shear direction and shear plane normal, respectively.

The shear rates on the active slip systems are evaluated using the Orowan equation

$$\dot{\gamma}^\alpha = \rho^\alpha b v^\alpha, \tag{42}$$

where the total dislocation density $\rho^\alpha$ is composted of the statistically stored dislocation (SSD) density $\rho_{SSD}^\alpha$ and GND density $\rho_{GND}^\alpha$, $\rho^\alpha = \rho_{SSD}^\alpha + \rho_{GND}^\alpha$. The dislocation velocity $v^\alpha$ on slip system $\alpha$ is specified constitutively as (Wong et al., 2016)

$$v^\alpha = \begin{cases} v_0 \exp\left\{ -\frac{Q_s}{k_B \theta}\left[1 - \left(\frac{\tau_{eff}^\alpha}{\tau_P}\right)^p\right]^q \right\} \mathrm{sign}(\tau^\alpha - \tau_{back}^\alpha) & \tau_{eff}^\alpha > 0 \\ 0 & \tau_{eff}^\alpha \leq 0 \end{cases}, \tag{43}$$

where $v_0$ is the dislocation glide velocity pre-factor, $Q_s$ is the activation energy, $k_B$ is the Boltzmann constant, $\tau_{\text{eff}}^\alpha$ is the effective resolved shear stress on slip system $\alpha$, $\tau_p$ is the Peierls stress, and the parameters $p$ and $q$ control the glide resistance profile.

We consider a particle reinforced material where the effective resolved shear stress $\tau_{\text{eff}}^\alpha$ is assumed as

$$\tau_{\text{eff}}^\alpha = \left| \tau^\alpha - \tau_{\text{back}}^\alpha \right| - \tau_{\text{pass}}^\alpha - \tau_{\text{orowan}}^\alpha - \tau_{\text{load}}^\alpha, \tag{44}$$

where $\tau_{\text{back}}^\alpha$ is the back stress induced by GNDs, $\tau_{\text{pass}}^\alpha$ is the forest dislocation passing stress, $\tau_{\text{orowan}}^\alpha$ is the Orowan stress required for particle bypassing, and $\tau_{\text{load}}^\alpha$ represents the load transferred to reinforcing particles.

The forest dislocation passing stress is given by

$$\tau_{\text{pass}}^\alpha = \kappa G b \sqrt{\sum_\beta \chi^{\alpha\beta} \rho^\beta}, \tag{45}$$

with the slip system interaction matrix $\chi^{\alpha\beta}$ and the Taylor coefficient $\kappa$.

The reinforcement effect resulting from dislocations bypassing particles is described by the Orowan stress (Fang et al., 2019),

$$\tau_{\text{orowan}}^\alpha = \frac{3Gb}{2L}, \tag{46}$$

where $L = r_p \sqrt{2\pi/3 f_p}$ denotes the average particle spacing, where $r_p$ is the particle radius and $f_p$ is the particle volume fraction.

Based on the shear-lag model for composite materials (Ryu et al., 2003), the expression for the load transfer stress contributed directly by load-bearing particles can be expressed as follows

$$\tau_{\text{load}}^\alpha = \sigma_{\text{int}} f_p \left(1 + \vartheta/2\right), \tag{47}$$

where $\sigma_{\text{int}}$ represents the intrinsic strengthening stress excluding the load-bearing contribution of particles, and $\vartheta$ is the effective aspect ratio of the particle, considered to be 1.0.

The evolution of SSD density includes both multiplication and annihilation processes of dislocations

$$\dot{\rho}_{\text{SSD}}^\alpha = \frac{\sqrt{\rho_f^\alpha}}{k_{\text{mult}}} \frac{|\dot{\gamma}^\alpha|}{b} - 2 d_{\text{anni}} \rho_{\text{SSD}}^\alpha \frac{|\dot{\gamma}^\alpha|}{b}, \tag{48}$$

where $\rho_f^\alpha$ is the forest dislocation density, the parameter $k_{\text{mult}}$ measures the characteristic slip distance of a dislocation in multiples of forest dislocation spacings, and $d_{\text{anni}}$ is the critical annihilation distance for dislocations. The forest dislocation density on the $\alpha$ slip system is expressed as $\rho_f^\alpha = \sum_\beta \rho^\beta |\mathbf{n}^\alpha \cdot \mathbf{t}^\beta|$, with the edge dislocation line direction $\mathbf{t}^\beta$ given by $\mathbf{t}^\beta = \mathbf{m}^\beta \times \mathbf{n}^\beta$.

The evolution rates of the GND densities $\dot{\rho}_{\text{GND}}^\alpha$ derive from the spatial derivatives of the corresponding slip rates (Cheong et al., 2005; Ma et al., 2006)

$$\dot{\rho}_{\text{GND}}^\alpha = \frac{1}{b} \left\| \nabla \times \left( \dot{\gamma}^\alpha \mathbf{n}^\alpha \cdot \mathbf{F}_p^T \right) \right\|. \tag{49}$$

In this work, we relate the back stress $\tau_{\text{back}}^\alpha$ to gradient of the GND density [38]. Additionally, considering HDI hardening by heterostructure, we introduce heterogeneity coefficient $\delta_h$

$$\tau_{\text{back}}^\alpha = \frac{(1+\delta_h)DGb}{2\pi} \left( \frac{\nabla \rho_{\text{GND}}^\alpha}{2\rho^\alpha(1-\upsilon)} \cdot \mathbf{m}^\alpha + \frac{\nabla \rho_{\text{GND}}^\alpha}{2\rho^\alpha} \cdot \mathbf{t}^\alpha \right). \tag{50}$$

In this expression, $D$ is a constant, taken as 3.0, $\upsilon$ is Poisson's ratio, and $k_h$ is the heterogeneity parameter.

Back stress is enhanced when the stress of strong dislocation pile-ups in large grains are not balanced by corresponding forward stress from smaller neignboring grains, in other words, back stress is boosted by grain heterogeneitiy. We describe this effect in terms of a heterogeneity coefficient $\delta_h$ that quanties the uniformity of grain size over a specified region of radius $r_h$ in terms of the weighted standard deviation

$$\delta_h = \sqrt{\sum_i f_i \left( \frac{d_i - \bar{d}}{d_i} \right)^2} \quad \text{with} \quad \bar{d} = \sum_i f_i d_i, \tag{51}$$

where $d_i$ represents the diameter of the $i$-th grain within the region, and its corresponding volume fraction $f_i$ can be calculated by the ratio of its area $s_i$ to the region's area $f_i = s_i / \pi r_h^2$. Numerical calculation of the heterogeneity coefficient takes into account the periodicity of the model.

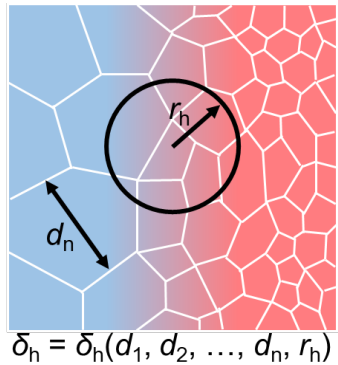

$\delta_h = \delta_h(d_1, d_2, \ldots, d_n, r_h)$

**Fig. 2.** The heterogeneity coefficient in a periodic model, which depends solely on the grain size $d$ within the influence zone (radius $r_h$).

In the present study, PBCs are employed to eliminate spurious boundary effects and to ensure that the simulated response reflects the intrinsic behavior of heterogeneous bulk microstructures. Moreover, the model used in this work is a nonlocal model involving the computation of curl and gradient for internal variables; in fact, due to the use of the back stress the constitutive model implies the evaluation of second-order strain gradients. Due to the difficulty of directly computing these higher-order gradients using conventional finite element methods, the Fourier transform is employed to achieve rapid and efficient solutions in Fourier space. By transforming the problem into the Fourier domain, the spatial derivatives (gradient and curl) are converted into simple algebraic operations. This significantly reduces the computational complexity, as the Fourier transform allows for the computationally efficient, parallelized computation of these operators. Moreover, the periodicity inherent in Fourier space is particularly suitable for periodic boundary conditions (PBCs) as considered here. This further enhances numerical stability and accuracy, making the method particularly efficient for solving nonlocal models with large-scale internal variable interactions.

However, problems may arise due to the inherent discontinuities of slip at grain boundaries, which may give rise to numerical artefacts and mesh sensitivity. To mitigate mesh sensitivity issues in the evaluation of GND densities, a spectral regularization is applied when computing the curl operator in the Fourier domain. Specifically, a high-frequency filter of the form $i\mathbf{k}/1+(\eta\lambda)^2 k^2$ is introduced to smooth the solution and enhance numerical stability

$$\nabla \times f(x) = \mathcal{F}^{-1}\left[\frac{i\mathbf{k}}{1+(\eta\lambda)^2 k^2} \times \overline{f}(\mathbf{k})\right], \qquad (52)$$

with the filter coefficient $\eta$ and grid spacing $\lambda$. The function $\overline{f}(\mathbf{k})$ represents the Fourier transform of the function $f(x)$, given as $\overline{f}(\mathbf{k}) = \mathcal{F}[f(x)]$, $\mathbf{k}$ is the frequency vector in Fourier space, $\mathcal{F}^{-1}[\cdot]$ is the inverse Fourier transform, and $k$ denotes the magnitude of the vector $\mathbf{k}$.

It is worth noting that this regularization is only applied to the curl calculation, while the gradient operator $i\mathbf{k}$ is retained without a filter

$$\nabla f(x) = \mathcal{F}^{-1}\left[i\mathbf{k}\overline{f}(\mathbf{k})\right]. \qquad (53)$$

## 3. Parameterization and validation of the constitutive model

The constitutive model is implemented using the Düsseldorf Advanced Material Simulation Kit, DAMASK (Roters et al., 2019). The Fast Fourier Transform (FFT) method is employed to solve boundary value problems, thereby restricting it to PBCs. In this work, uniaxial tension is applied in the X direction, with the strain rate set to 0.001/s. The imposed constraints on strain rate and stress are given by (Shanthraj et al., 2019)

$$\dot{\bar{\mathbf{F}}} = \begin{bmatrix} 1 & 0 & 0 \\ 0 & * & 0 \\ 0 & 0 & * \end{bmatrix} \times 10^{-3} \text{ s}^{-1} \text{ and } \bar{\mathbf{P}} = \begin{bmatrix} * & * & * \\ * & 0 & * \\ * & * & 0 \end{bmatrix} \text{Pa}, \qquad (54)$$

'∗' indicates that the corresponding degree of freedom is unconstrained. PBCs are applied in all directions. The macroscopic strain is prescribed in the loading direction, while the macroscopic stresses in the transverse and thickness directions are constrained to zero. Although the model is periodic along the thickness direction and thus has no free surfaces, the overall deformation state corresponds to a plane-stress-like periodic condition.

As a typical example of a HS material, this study conducts numerical simulations of the deformation behavior based on the experimental microstructure of a HS Al-matrix composite as reported by Nie et al. (Nie et al., 2023). The microstructural grain morphology obtained from experimental characterization (Fig. 3(a)), serves as the basis for constructing a representative volume element (RVE) model (Fig. 3(b)). The RVE considered in this study has dimensions of $50 \times 50 \times 0.25$ μm³. The model comprises fine-grained (FG) and CG regions, occupying 80% and 20% of the total volume, respectively. Experimental observations reveal that AlN nanoparticles are distributed within the FG regions only, therefore particle strengthening is consiered exclusively in these regions. Due to the extrusion process used for manufacuring, the material exhibits a fiber texture <111>//X along a specific direction, as illustrated by the inverse pole figure (IPF) map (Fig. 3(c)). To ensure consistency with experimental findings, the same fiber texture is implemented in the simulation (Fig. 3(d)).

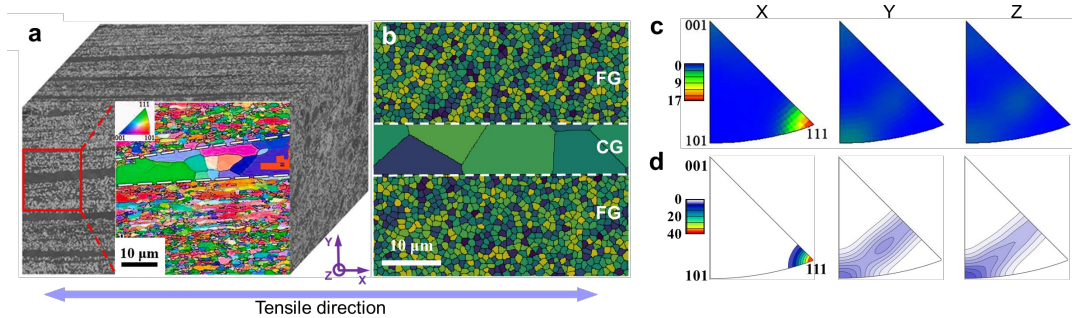

**Fig. 3.** (a) Microstructural features and grain morphology of the HS material as reported by Nie et al. (Nie et al., 2023). (b) Geometric model used for simulation. (c) Experimental fiber texture observed in Ref. (Nie et al., 2023). (d) Grain orientation in the simulation is set to match the experimental fiber texture.

To examine the sensitivity of the heterogeneity coefficient and the back stress to the choice of the heterogeneous influence zone, we present their spatial distributions for different values of the radius (Fig. 4). Notably, the radius of the heterogeneous influence zone determines the back stress distribution, which in turn affects the internal stress state of the material. Specifically, results are shown for $r_h$ = 1.25 μm, 2.5 μm, and 3.75 μm. $r_h$ = 2.5 μm is adopted in the present work, and the rationale for this choice is discussed in Section 4.2. This choice ensures that the heterogeneous influence zone is larger than the average grain size in the ultrafine domain but smaller than that in the CG domain, thereby allowing the heterogeneity coefficient to capture interfacial interactions across length scales without being overly localized or excessively homogenized. The characteristic width of the crack $l_c$ is chosen to be approximately twice the finite element size, in line with established practice in PF simulations of damage in crystalline materials (Diehl et al., 2017; Li et al., 2022; Shanthraj et al., 2016). The remaining parameters used in the constitutive model are listed in Table 1. These parameters are determined based on a combination of experimental data and established literature, with key values (such as Peierls stress, hardening coefficient, and intrinsic strengthening stress from precipitates) identified through calibration against CG and HS tensile responses. The six interaction types among the twelve slip systems of FCC (Kubin et al., 2008), and the value of interaction strength is shown in Table 2. We note that the plastic dissipation inequality, Eq. (26), in conjunction with the dislocation defect energy, Eq. (37), imposes constraints on the parameters of the crystal plasticity model. These constraints have been taken into account following the approach of Wu and Zaiser (2022). Further details on the numerical validation of damage evolution, including mesh sensitivity and the effects of PBCs, are provided in Appendix C.

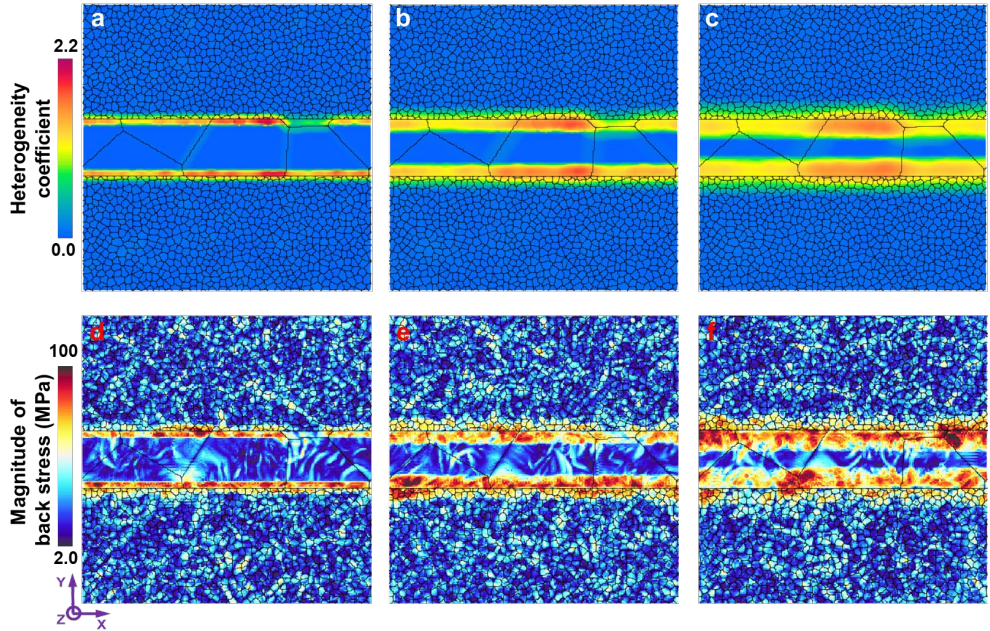

**Fig. 4.** Distribution of the heterogeneity coefficient and the magnitude of back stress at 10% strain for different radii of the heterogeneous influence zone: (a, d) $r_h$ = 1.25 μm, (b, e) $r_h$ =

2.5 μm, and (c, f) $r_h$ = 3.75 μm.

**Table 1** Material parameters used in the CP-PF model

| Module | Symbol | Description | Value and unit | Source |
|---|---|---|---|---|
| CP | $C_{11}$ | Elastic moduli | 106.75 GPa | (Kords and Raabe, 2013) |
| | $C_{11}$ | | | |
| | $C_{12}$ | | 60.41 GPa | |
| | $C_{44}$ | | 28.34 GPa | |
| | $\rho^\alpha$ | Initial dislocation density | $1.0 \times 10^{10}$ m/m$^3$ | (Kords and Raabe, 2013) |
| | $b$ | Magnitude of the Burgers vector | $2.86 \times 10^{-10}$ m | (Kords and Raabe, 2013) |
| | $v_0$ | Dislocation glide velocity pre-factor | $1.0 \times 10^{-4}$ m/s | (Roters et al., 2019) |
| | $Q_s$ | Activation energy | $2.5 \times 10^{-19}$ J | (Kords and Raabe, 2013) |
| | $p$, $q$ | Energy barrier profile constants | 1.0 | (Roters et al., 2019) |
| | $\tau_P$ | Peierls stress | 50 MPa | Fit CG |
| | $k_{mult}$ | Hardening parameter | 40.0 | Fit CG |
| | $\kappa$ | Taylor coefficient | 0.3 | (Cheong et al., 2005) |
| | $d_{anni}$ | Annihilation distance of dislocations | $1.87 \times 10^{-9}$ m | (Roters et al., 2019) |
| | $f_p$ | Particle volume fraction | 0.12 | (Nie et al., 2023) |
| | $r_p$ | Particle radius | $9.53 \times 10^{-8}$ m | (Nie et al., 2023) |
| | $\sigma_{int}$ | Intrinsic strengthening stress | 300 MPa | Fit HS1 |
| | $r_h$ | Heterogeneous influence zone | $2.5 \times 10^{-6}$ m | |
| | $k_h$ | Heterogeneity parameter | 3.0 | |
| | $\eta$ | Filter coefficient | 8.0 | Appendix B |
| PF | $\psi_{cr, CG}$ | Critical energy density for CG | $1.5 \times 10^6$ J/m$^2$ | Fit CG |
| | $\psi_{cr, FG}$ | Critical energy density for FG | $3.0 \times 10^6$ J/m$^2$ | Fit HS1 |
| | $g_c$ | Critical energy release rate | 1.0 J/m$^2$ | |
| | $l_c$ | Characteristic width of the crack | $0.5 \times 10^{-6}$ m | |
| | $M_f$ | Mobility | 2.0 m$^3$/(Js) | |

**Table 2** The interaction types and strength coefficients of FFC crystals (Kubin et al., 2008)

| Interaction type | Interaction coefficient |
| --- | --- |
| Self interaction | 0.122 |
| Coplanar interaction | 0.122 |
| Collinear interaction | 0.625 |
| Hirth lock | 0.07 |
| Glissile junction | 0.137 |
| Lomer lock | 0.122 |

Introducing a high-frequency filter in the Fourier-based computation of the curl operator significantly mitigates mesh sensitivity in the strain gradient constitutive model, as discussed in Appendix B. The effectiveness and necessity of this spectral regularization are validated through a detailed analysis of stress-strain curves and of the distribution of internal variables.

Following the optimization of mesh sensitivity, a comparative study between experimental and simulated results is conducted, focusing on stress-strain response, crack propagation behavior, and dislocation distribution. This comparison underscores the critical role of the heterogeneity coefficient in the model. Fig. 5(a) presents a comparison of uniaxial tensile stress-strain curves obtained from experiments and simulations. For completeness, the tensile response of the FG structure is also included in Fig. 5(a), providing a direct reference for evaluating the relative deformation behaviors of the FG, CG, and HS configurations. The simulation results exhibit well agreement with the experimental data for both uniform CG and HS laminates. Damage initiation is observed at triple junctions, highlighting these regions as inherent weak points in the material.

To elucidate the additional strengthening effect of back stress, Fig. 5(b) also includes a tensile process calculated without considering back stress. It is important to clarify that the term "without back stress" refers to setting the back stress to zero while maintaining a nonzero GND density (with strain gradient). The ROM curve in Fig. 5(b) is obtained by separately calculating the CG and FG responses and averaging them by volume fraction, thereby explicitly incorporating the FG contribution. The close agreement between the ROM curve and the HS simulation without back stress indicates that strain gradients alone do not generate additional strengthening; the extra resistance arises from the back stress associated with hetero-deformation.

The contribution of back stress is most prominent during the yield stage, where it increases the material strength by approximately 18 MPa, with minimal influence on subsequent strain hardening. Furthermore, Fig. 5(c) provides a quantitative decomposition of the strengthening contributions at 10% strain, where the load transfer in FG from precipitates (137.0 MPa), Orowan strengthening (76.3 MPa), back stress (18.0 MPa), and passing stress (177.9 MPa) are separately evaluated. In addition, the effect of the critical energy density $\psi_{cr, FG}$ on the damage behavior is analyzed, as shown in Fig. 5(d). This parameter primarily affects the fracture strain, where a lower value promotes earlier damage initiation. Even when the critical energy density is identical for the FG and CG regions, crack still preferentially propagates through the FG region.

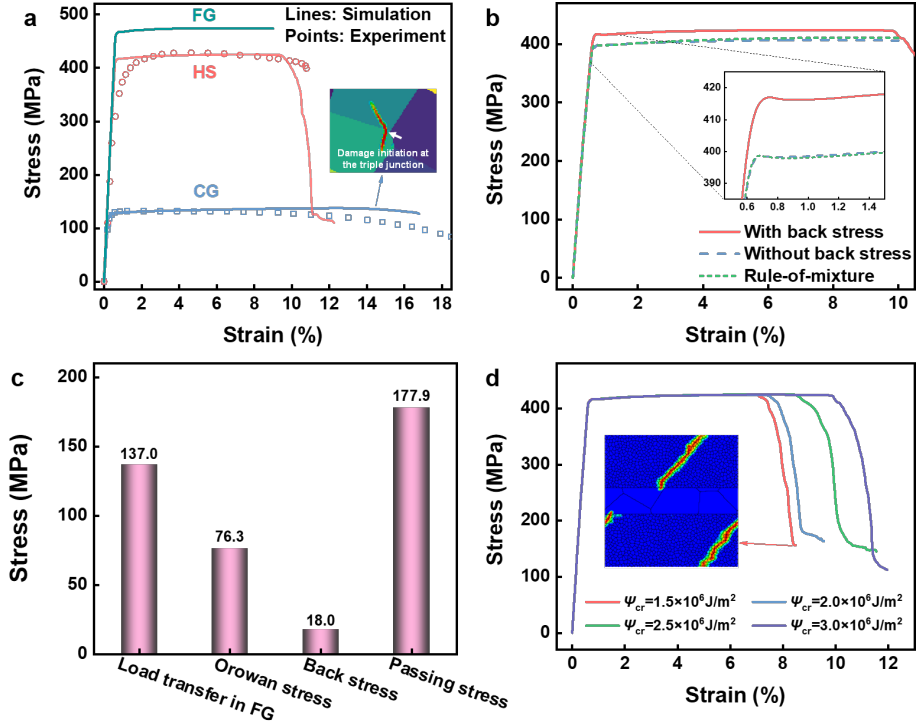

**Fig. 5.** (a) Comparison of experimental (Nie et al., 2023) and simulated uniaxial tensile stress-strain curves for CG, FG and HS materials. (b) Simulation results highlight the contribution of back stress by comparing stress-strain responses with and without back stress and demonstrating the additional contribution of heterogeneity using the rule-of-mixture without back stress. (c) Quantitative contributions of different strengthening mechanisms at 10% strain. (d) Influence of the critical energy density $\psi_{cr, FG}$ on the stress-strain response and fracture strain.

## 4. Deformation mechanisms of HS laminates

### 4.1 Microscale damage behavior

Fig. 6 compares the simulated and experimental crack morphologies as well as the evolution of slip bands at different strain levels. Figs. 6(a1)-(a3) illustrate the simulated crack evolution as the applied strain increases from 11% to 14%. Microcracks initiate from the HIs and extend toward the FG regions. Once the cracks traverse the FG region, their propagation becomes significantly impeded, leading to crack arrest at the FG-CG interface. A similar crack arrest phenomenon at the heterogeneous interface is also observed experimentally, as shown in Fig. 6(b) (Fan et al., 2025). With further straining up to 14%, the arrested cracks eventually penetrate through the CG region, coalescing into a dominant through-thickness crack.

The corresponding evolution of plastic slip, shown in Figs. 6(c1)-(c3), reveals a much higher density of slip lines and slip bands in the CG regions compared with the FG regions. indicating a reduced degree of slip concentration. (Note that overall slip activity must be compatible, hence the fewer slip bands in the FG region must carry larger local strains). This observation is consistent with the experimental results in Fig. 6(d) (Nie et al., 2023). Only after a crack has crossed the FG region, the ensuing stress concentration at the crack tip induces pronounced slip localization also in the CG region, within slip bands that are typically anchored at the crack tips at both interfaces. The final crack path through the CG region closely follows the trajectory of these concentrated slip bands.

Crack arrest at the HI is thus governed by the initially high plastic activity and low degree of strain localization in the CG region, where the enhanced deformation compatibility blunts the crack tip and suppresses further propagation. This arrested state does not restore the stress lost during FG cracking; instead, it creates a stable configuration in which the CG region can continue to accommodate substantial plastic strain through distributed deformation, delaying the onset of catastrophic, through-thickness failure. Crack propagation across the CG region becomes only possible in a second step, once the stress concentration at the HI has caused more pronounced strain localization and concomitant damage in a slip band crossing the CG region, see Figs. 6 (a3, a4) and (c3, c4).

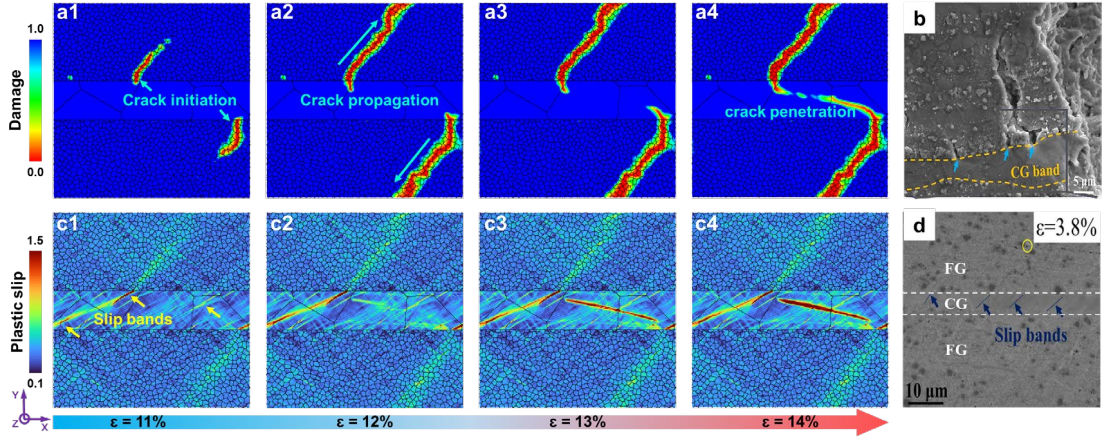

**Fig. 6.** Crack morphologies at different strain levels: (a1) 11%, (a2) 12%, (a3) 13%, and (a4) 14%. (b) Experimental observation showing that cracks initiate from the CG-FG interface and initially propagate into the FG region (Fan et al., 2025). Plastic slip distributions at different strain levels: (c1) 11%, (c2) 12%, (c3) 13%, and (c4) 14%. (d) Experimental observation revealing the formation of numerous slip bands within the CG region (Nie et al., 2023).

To elucidate the mechanism of damage initiation, the evolution of average defect energy with increasing strain for the CG and FG regions is plotted in Fig. 7(a). After a short initial transient, the defect energy in the FG regions exceeds that in the CG region, leading to preferential damage initiation within the FG region. The spatial distribution of defect energy at 10% strain shown in Fig. 7(b) demonstrates that the defect energy in the FG regions is enhanced near the FG-CG between the FG-CG interface, which explains why these near-interface regions act as preferential crack nucleation sites. Because the CG region maintains a consistently lower defect energy, it provides a higher resistance to defect accumulation. This intrinsic defect-energy contrast causes cracks initiated in the FG region to slow down and become arrested upon entering the lower-energy CG region.

To further assess the influence of strain gradient effects we compared the plastic slip and crack initiation patterns obtained with and without considering GNDs. When GNDs are included, pronounced slip bands appear within the CG regions, whereas little slip localization is observed in the absence of GNDs, as shown in Figs. 7(c) and (d). The effect of GNDs on the mechanical response manifests in two ways: (i) strengthening through dislocation forest interactions, and (ii) generation of back stress, which enhances long-range internal stress gradients. Consequently, the inclusion of GNDs amplifies strain heterogeneity and promotes the formation of slip bands. This results in different crack initiation patterns as seen by comparing Figs. 7(e) and (f). When GNDs are accounted for, cracks preferentially initiate from the HIs; in contrast, without GNDs, cracks nucleate randomly within the FG regions.

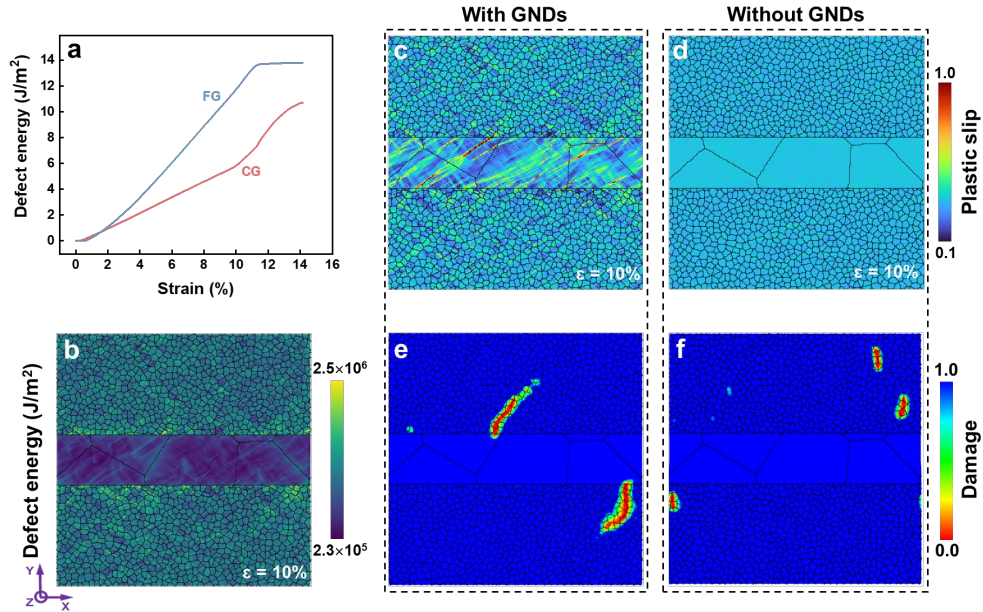

**Fig. 7.** (a) Evolution of average defect energy in the CG and FG regions of the HS material during deformation. (b) Spatial distribution of defect energy at 10% strain (undamaged state). Plastic slip distribution at 10% strain: (c) with geometrically necessary dislocations (GNDs) considered and (d) without GNDs. Crack initiation sites: (e) with GNDs considered and (f) without GNDs.

## 4.2 Accumulation of dislocations

Figs. 8(a)-(c) show the GND distributions within the HS material at different deformation stages. At the early stage of deformation (1%), plasticity initiates preferentially in the CG regions. The strength mismatch across the HIs leads to pronounced strain incompatibility, necessitating the accumulation of GNDs to accommodate the non-uniform strain. Thus, GNDs accumulate in near-interface regions of coarse grains (Fig. 8(a)). As the strain increases to 10%, a pronounced concentration of GNDs remains near the HIs and along the CG boundaries (Fig. 8(b)). Experimental kernel average misorientation (KAM) maps, shown in Figs. 8(d, e), confirm this trend by showing that local misorientations within the CG are enhanced near the HI (Shukla et al., 2018).

Crack propagation into the CG regions is preceded by the formation of localized slip bands crossing these regions. These bands emanate from stress concentrations at the locations where the crack connects to the HI (see Fig. 6(c3). Along these slip bands, a noticeable increase in GND density is observed (Fig. 8(c)). The crack follows this zone of increased GND density (Fig. 6(a4, c4), Fig. 8(c)). This observation is consistent with EBSD data shown in Fig. 8(f) (Dang et al., 2025), indicating ductile failure features with a high density of GNDs in the fracture zone. These observations illustrate the strong, GND mediated coupling between plastic deformation and damage evolution in the CG region.

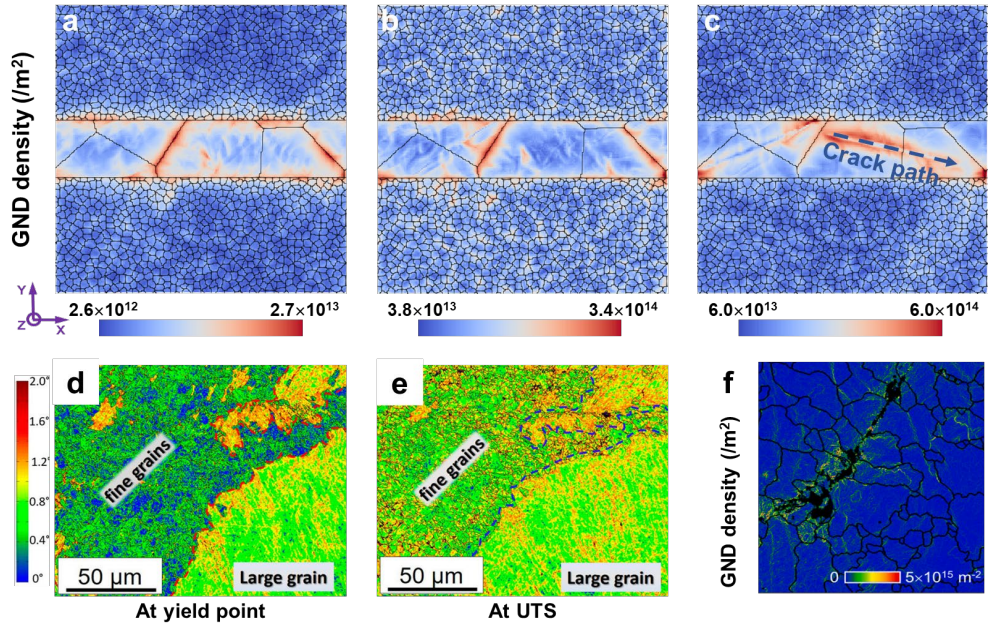

**Fig. 8.** Comparison of GND density and kernel average misorientation (KAM) distributions at different deformation stages. Simulated GND density distributions at strains of (a) 1%, (b) 10%, and (c) 14% (damaged state). Experimentally measured KAM maps: (d) just below the yield point and (e) at the ultimate tensile strength (UTS) (Shukla et al., 2018). (f) Experimentally observed GND accumulation along the crack path (Dang et al., 2025).

Fig. 9 illustrates the dislocation distribution within the HS material at an early stage of deformation (0.5% strain). As plasticity initiates in the CG regions, the total dislocation density within the IAZs (indicated by the dashed region) becomes significantly higher than that in the surrounding areas, as shown in Fig. 9(a). The active slip systems in these regions are more easily triggered, promoting rapid dislocation multiplication and accumulation. The stress triaxiality distribution indicates that the IAZs experience a more complex multiaxial stress state induced by back stress, with local triaxiality values exceeding 0.5 (Fig. 9(b)).

Experimental observations confirm the relatively high dislocation density within the IAZs, exhibiting a spatial distribution consistent with the simulation results (Fig. 9(c)) (Xia et al., 2025). When the contribution of GNDs is not considered, the dislocation density within the CG region appears uniformly distributed at the same strain level, and the experimentally observed heterogeneous features of the IAZs cannot be reproduced (Fig. 9(d)). Moreover, the experiments demonstrate a complex multiaxial stress state within these zones, which facilitates the activation of additional slip systems even with low Schmid factors (Xia et al., 2025). The experimentally determined width of the interface-affected zone is approximately 3 μm, which agrees well with the influence zone radius $r_h$ = 2.5 μm as defined in Table 1.

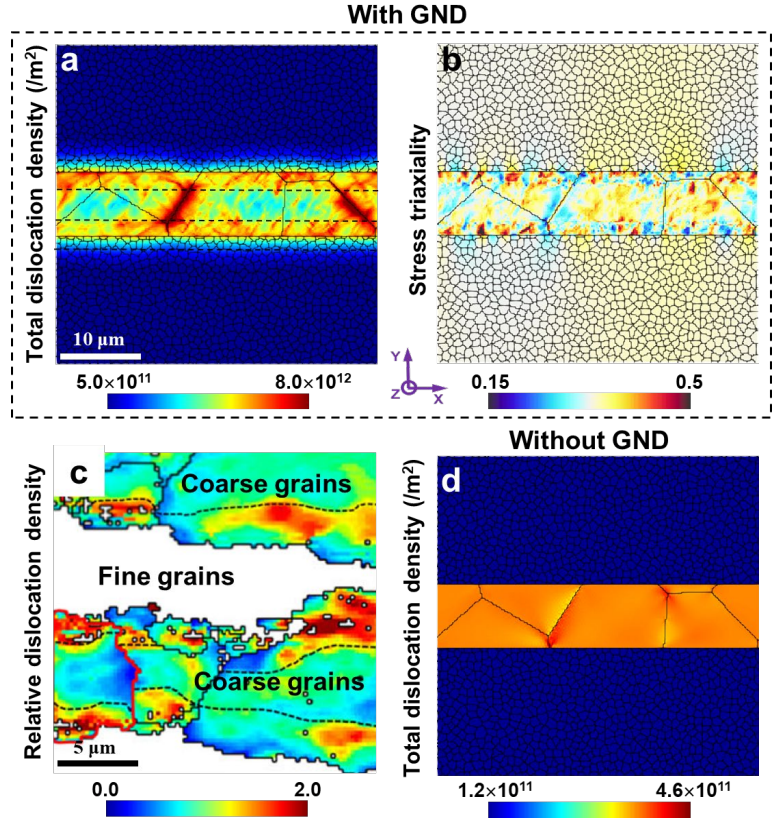

**Fig. 9.** Total dislocation density and stress triaxiality at 0.5% strain: (a) simulated total dislocation density, (b) stress triaxiality distribution, (c) experimentally characterized relative

dislocation density (Xia et al., 2025), and (d) simulated total dislocation density without considering GNDs.

## 5. Microstructure-mechanical properties relation

### 5.1 Stress-strain response in uniaxial tension

We constructed several model geometries to investigate the influence of microstructure configuration on strength and ductility. By keeping the proportion of CG regions constant, we varied the number of CG layers in the RVE, thereby altering the corresponding layer thicknesses (10 μm, 5 μm, and 3.3 μm, respectively). The different configurations of HS laminates are labeled as HS1, HS2, and HS3, representing 1-layer, 2-layer, and 3-layer CG regions, respectively, as shown in Fig. 10. This coefficient characterizes the heterogeneity within a specific area, predominantly concentrated near the interface between CG and FG regions, and gradually decreases as the grain size of the CG regions reduces. The HS models are all meshed with a size of 200×200×1 grids and have geometric dimensions of 50×50×0.25 μm$^3$.

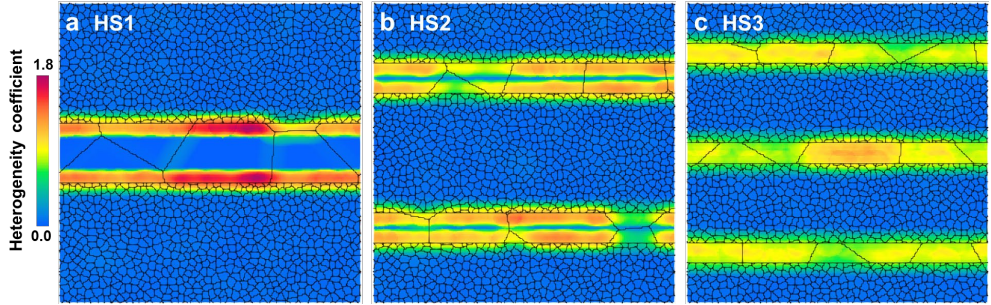

**Fig. 10.** Grain morphology and heterogeneity coefficient $\delta_h$ distribution of the three HS models: (a) HS1, (b) HS2, (c) HS3.

This section investigates the relationship between configuration and ductility by varying the layer thickness while maintaining a constant CG phase volume fraction. In this work, ductility refers to the material's load-bearing capacity in the post-initiation regime, i.e., after the crack has propagated through the FG layer and entered the CG layer, where the mechanical response enters the second stage of degradation. Uniaxial tensile simulations are conducted for the three configurations (Fig. 10), and their stress-strain curves are presented in Fig. 11(a). Damage initiation occurs at approximately 10% strain for all three HS materials. Rapid crack propagation within the FG regions leads to an initial sharp drop in stress. With further straining, the mechanical response of all HS materials exhibits a plateau at similar strain levels, although the stress level of the plateau differs among configurations: HS1 exhibits a lower plateau stress, while HS2 and HS3 display comparable plateau levels.

Given the inherent randomness of damage evolution in polycrystalline materials, four sets of simulations with different orientations are performed for each configuration. The load-bearing stresses at 11.6% strain are extracted for analysis, as shown in Fig. 11(b). At this strain level, the average stress is 117.7 MPa for HS1, 154.9 MPa for HS2, and 156.7 MPa for HS3. These results indicate that appropriately reducing the CG layer thickness can enhance ductility, but excessively thinning the CG layer facilitates through-thickness crack penetration, thereby limiting further performance improvement. This behavior suggests a trade-off between delaying damage initiation and maintaining post-initiation mechanical stability, consistent with experimental observations that indicate an optimal layer thickness for HS laminates (Huang et al., 2018).

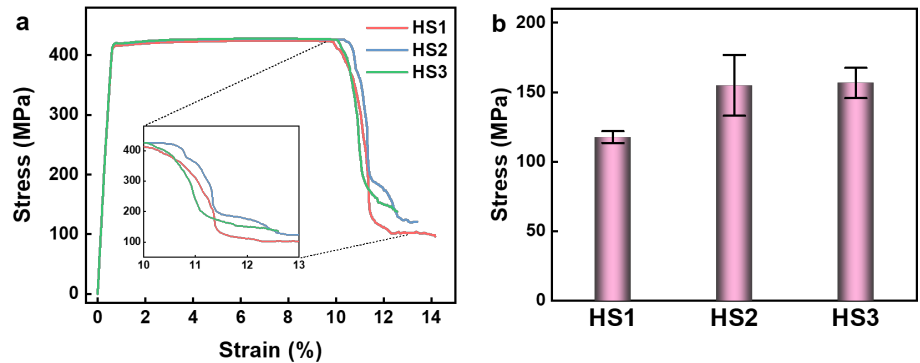

**Fig. 11.** (a) Stress-strain curves for different material configurations (HS1, HS2, HS3). (b) Load-bearing stresses at 11.6% strain for different configurations, based on four sets of simulations with different crystallographic orientations for each configuration.

**5.2 Microstructure evolution**

The mechanisms underlying strength and ductility can be further elucidated by analyzing crack morphologies at different stages of deformation, as shown in Fig. 12. During the strain increase from 11% to 12.6%, cracks in HS1 initiate and develop exclusively within the FG regions (Figs. 12(a1)-(a3)). In HS2, the increased density of HIs leads to shorter crack paths before they are obstructed by HIs. However, due to the finite thickness of the CG layers, cracks eventually penetrate through the CG regions when the strain reaches 12.6% (Figs. 12(b1)-(b3)). For HS3, although the number of obstructing regions is even higher, the significantly reduced thickness of the CG layers results in crack penetration through the CG region already at 12% strain, making it the earliest among the three configurations to exhibit through-thickness cracking (Figs. 12(c1)-(c3)). With an increasing number of CG layers, the probability of crack obstruction rises, leading to shorter crack lengths within the FG regions. Consequently, increasing the number of CG layers enhances the probability of crack arrest, thereby reducing the crack length within the FG regions and improving the overall ductility of the material. These results indicate that increasing the number of HIs by reducing the CG layer thickness increases the number of obstacles to crack propagation; however, excessive thinning of the CG layers significantly weakens their capacity to impede crack propagation. Thus, we observe a clear trade-off between increasing number and decreasing strength of obstacles, resulting in an intermediate optimum configuration.

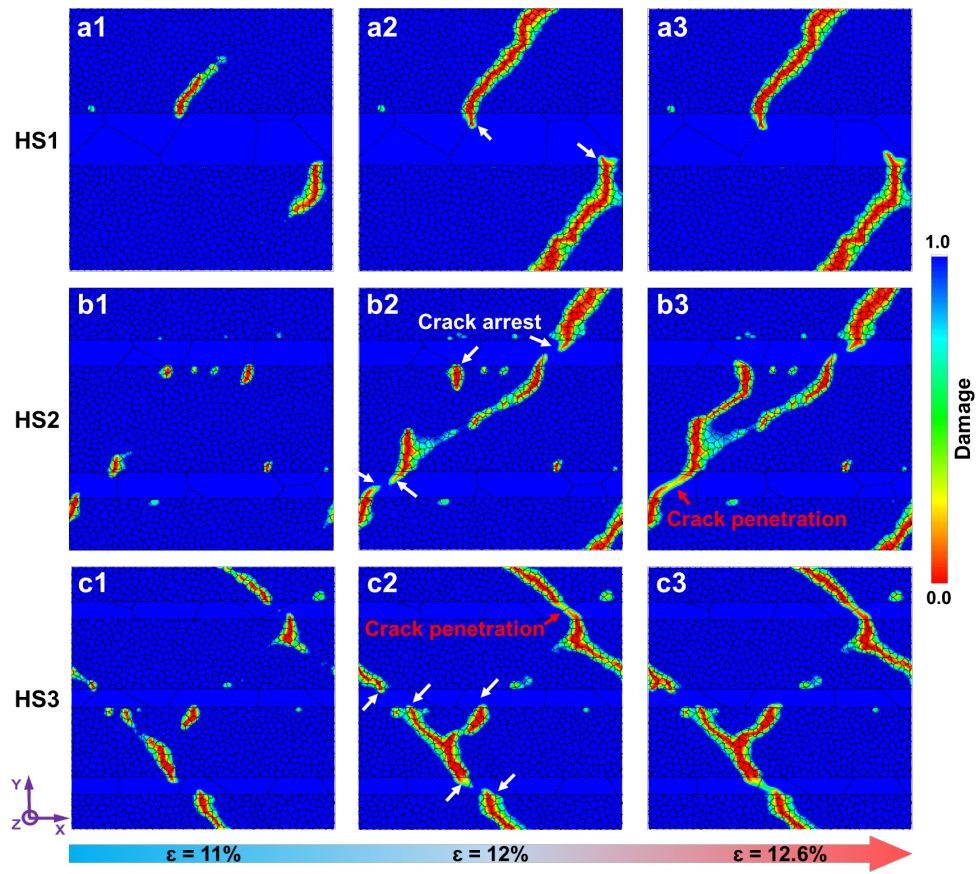

**Fig. 12.** Crack propagation process for different configurations of composites: (a1-a3) HS1, (b1-b3) HS2, and (c1-c3) HS3, illustrating the progression of cracks through the FG and CG regions at different stages of deformation.

Therefore, an appropriate reduction in CG layer thickness can effectively hinder crack propagation and enhance ductility. However, when the CG layers become too thin, their crack-arresting capability is substantially weakened. The influence of CG layer thickness on the ductility of HS laminates is closely related to the relationship between the crack-tip plastic zone size (approximately 2.5 μm, corresponding to half the thickness of the CG layer in HS2) and the layer thickness, as illustrated in Fig. 13. When the CG layer thickness exceeds twice the width of the crack-tip plastic zone, cracks are less likely to penetrate the CG layers. Conversely, if the thickness falls below this critical threshold, crack penetration becomes more likely.

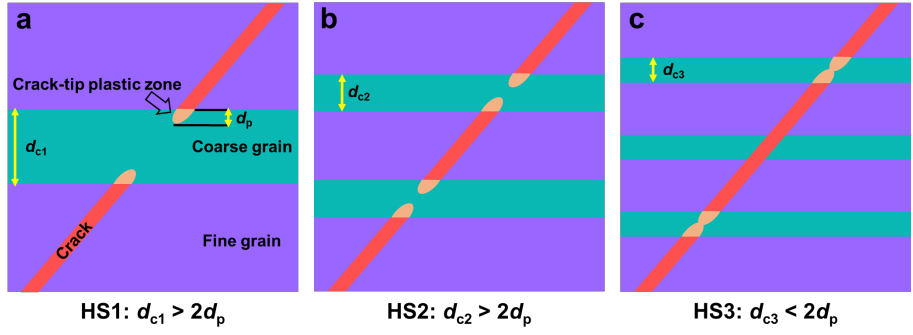

**Fig. 13.** Relationship between crack-tip plastic zone $d_p$ and CG layer thickness $d_c$ for three HS configurations: (a) HS1: $d_{c1} > 2d_p$, (b) HS2: $d_{c2} > 2d_p$ and (c) HS3: $d_{c3} < 2d_p$.

To analyze the deformation characteristics during the plastic stage for different configurations, the distributions of plastic slip (Figs. 14(a)-(c)) and GND density (Figs. 14(d)-(f)) at a strain of 10%—prior to damage initiation—are presented. Even in the absence of crack initiation, pronounced slip bands are already observed within the CG regions, which serve to accommodate and coordinate deformation. As the CG layer thickness decreases, the number of dispersed slip bands increases, which is beneficial for alleviating strain localization within the CG regions. Moreover, all HS materials exhibit pronounced GND accumulation along the HIs and certain CG boundaries, with the GND density at these interfaces increasing as the CG layer thickness increases. These observations indicate that the design of heterogeneous structures can significantly influence the microstructural evolution during the plastic deformation stage.

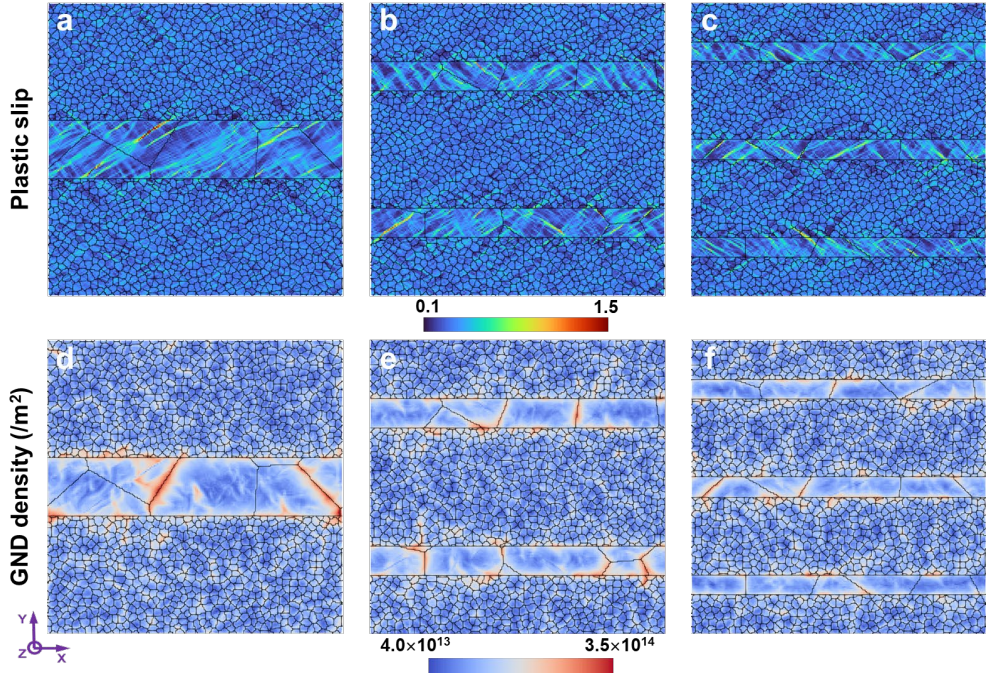

**Fig. 14.** Comparison of plastic slip and GND density distributions at 10% strain for different configurations: (a-c) plastic slip distributions for HS1, HS2, and HS3, respectively; (d-f)

corresponding GND density distributions for HS1, HS2, and HS3, respectively.

The crack initiation and propagation paths within the material are shown to be closely correlated with the distribution of defect energy, as illustrated in Fig. 15. Although the IAZs within the CG regions exhibit relatively high GND densities, the total dislocation density in the FG regions increases rapidly, surpassing that in the CG regions. Nevertheless, the presence of concentrated plastic slip bands within the CG regions locally elevates the defect energy, thereby promoting crack propagation along these plastically deformed zones. At a strain of 12.6%, the defect energy density along the crack paths in HS2 and HS3 is higher than in HS1, indicating an enhanced resistance to crack propagation in these configurations. These results further highlight the importance of employing a defect-energy-based ductile fracture model for accurately capturing the failure behavior of HS materials.

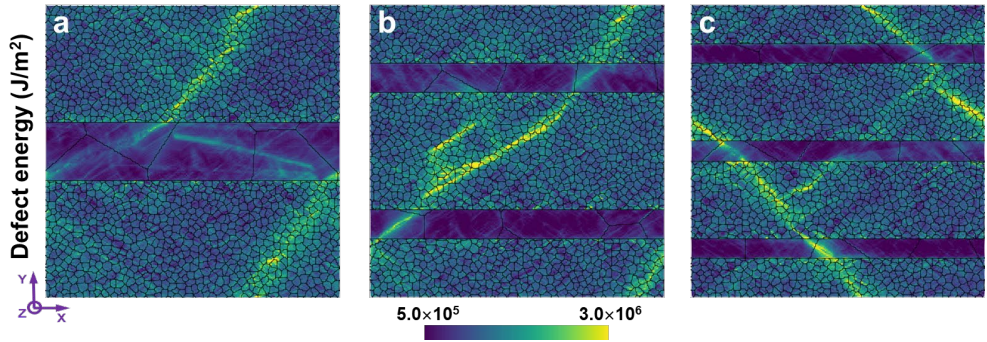

**Fig. 15.** Comparison of defect energy distributions at 12.6% strain for different configurations: (a) HS1, (b) HS2, and (c) HS3.

## 6. Conclusions

This work develops a coupled strain gradient CP and PF damage model to investigate the deformation and fracture mechanisms in HS laminates. The framework efficiently solves nonlocal gradient and curl operators using a Fourier-based spectral approach under PBCs. To address mesh sensitivity in GND density evaluation, spectral regularization is introduced specifically for the curl operator via a high-frequency filter.

The model successfully captures interface-induced back stress strengthening and quantitatively reproduces experimental stress-strain responses. The simulated plastic slip distribution shows strong agreement with experimental observations, accurately reflecting the strain localization near HIs. Strength enhancement is found to originate from two synergistic mechanisms: the accumulation of GNDs near interfaces driven by strain gradients, and the resulting back stress due to heterogeneous GND distributions.

In terms of fracture behavior, cracks are observed to initiate at HIs and propagate preferentially along FG regions before being arrested in CG zones. This sequential path enhances crack resistance while maintaining ductility. Moreover, thinning the CG layers improves crack-arrest capability, but further reduction does not result in additional gains in toughness.

While validated in HS laminates, the modeling framework is generally applicable to a broad range of HS materials with spatially varying microstructures. This study provides fundamental insight into the interplay between heterogeneity, plasticity, and damage, offering a predictive tool for designing and optimizing heterostructures with superior mechanical performance.

**Acknowledgments**

This work was supported by the National Natural Science Foundation of China (Grant numbers: 12222209, 12202423, 12192214, 52192591), and Sichuan Science and Technology Program (Grant numbers. 2024NSFCJQ0068, 2025HJRC0006). M.Z. also acknowledges support by DFG under grant no. 1Za 171/13-1, and MZ and XZ acknowledge support by the Sino-German Research Office under Exchange Grant No. M-0511.

**Data availability**

Data will be made available on request.

**Appendix A. Interface energy scaling**

Eq. (39) represents the surface energy of the discrete crack surface. According to the general approach by Cahn and Hilliard (1958), the surface energy density in a 1D domain consists of both a homogeneous and a gradient component

$$\rho_0 \psi_f = \psi_h(\varphi(x)) + \psi_g(\partial_x \varphi(x)), \quad \psi_g(\partial_x \varphi) = \psi_{g0} |\partial_x \varphi|^2, \tag{A1}$$

where $\psi_h$ represents the local surface energy, $\psi_g$ is used for the crack regularization, i.e., smoothing the crack surface, and $\psi_{g0}$ is a constant. The total crack surface energy density in the 1D domain can be expressed as

$$g_c = \int_{-\infty}^{\infty} \psi_f(x)\, \mathrm{d}x. \tag{A2}$$

In the equilibrium state $\delta_\varphi \psi_f = 0$, by combining the boundary conditions $\varphi(\pm\infty) = \phi_\pm$ and $\partial_x \varphi(-\infty) = 0$, one can write

$$\partial_\varphi \psi_h = \partial_x \partial_{\partial_x \varphi} \psi_g = 2\psi_{g0} \partial_x \partial_x \varphi. \tag{A3}$$

Multiplying both sides by $\partial_x \varphi$ and integrating yields

$$\int_{-\infty}^{x} 2\psi_{g0} \partial_\xi \partial_\xi \varphi \partial_\xi \varphi \, d\xi = \int_{-\infty}^{x} \partial_\xi \psi_g \left(\partial_x \varphi(\xi)\right) d\xi = \psi_g \left(\partial_x \varphi(x)\right),$$
$$\int_{-\infty}^{x} \partial_\varphi \psi_h \partial_\xi \varphi \, d\xi = \int_{-\infty}^{x} \partial_\xi \psi_h \left(\varphi(\xi)\right) d\xi = \int_{\varphi_-}^{\varphi(x)} \partial_\zeta \psi_h(\zeta) \, d\zeta = \Delta \psi_h(\varphi(x)), \tag{A4}$$

where $\Delta \psi_h(\varphi(x)) = \psi_h(\varphi(x)) - \psi_h(\varphi_-)$, based on Eqs. (A1) and (A4), the following equation can be established

$$\psi_g \left(\partial_x \varphi(x)\right) = \psi_{g0} \left|\partial_x \varphi\right|^2 = \Delta \psi_h(\varphi). \tag{A5}$$

Thus, the gradient term can be given as

$$\left|\partial_x \varphi\right| = \sqrt{\frac{\Delta \psi_h(\varphi)}{\psi_{g0}}}. \tag{A6}$$

By applying the differential transformation $dx = (1/\partial_x \varphi) d\varphi$, the following transformed form can be obtained

$$dx = \text{sign}(\partial_x \varphi) \sqrt{\frac{\psi_{g0}}{\Delta \psi_h(\varphi)}} \, d\varphi. \tag{A7}$$

By substituting Eq. (A7) into Eq. (A2) and combining with Eq. (A5), we derive

$$g_c = \kappa_{h0} \sqrt{\psi_{g0}}, \quad \kappa_{h0} = \text{sign}(\partial_x \varphi) \int_{\varphi_-}^{\varphi_+} \frac{\psi_h(\varphi) + \Delta \psi_h(\varphi)}{\sqrt{\Delta \psi_h(\varphi)}} \, d\varphi. \tag{A8}$$

Assuming that damage initiates at $x = 0$, and defining the characteristic width $l_c$, i.e., $\varphi(\pm l_c/2) \approx \varphi_\pm$, the following can be derived using a Taylor expansion:

$$\varphi(x) = \varphi(0) + \partial_x \varphi(0) x + \frac{1}{2} \partial_x^2 \varphi(0) x^2 + \cdots$$
$$\varphi_+ - \varphi_- = \partial_x \varphi(0) l_c - \frac{1}{24} \partial_x^3 \varphi(0) l_c^3 + \cdots \approx \partial_x \varphi(0) l_c. \tag{A9}$$

Substituting Eq. (A9) into Eq. (A6), we obtain

$$l_c = \kappa_{g0} \sqrt{\psi_{g0}}, \quad \kappa_{g0} = \frac{|\varphi_+ - \varphi_-|}{\sqrt{\Delta \psi_h(\varphi(0))}}. \tag{A10}$$

By combining Eqs. (A8) and (A10), $\psi_{g0}$ can be expressed as

$$\psi_{g0} = \frac{g_c l_c}{\kappa_{h0} \kappa_{g0}}. \tag{A11}$$

By considering the scaled expression for the homogeneous energy, $\psi_h(\varphi) = \psi_{h0} \varphi_h(\varphi)$,

with $\psi_{h0}$ constant and $\varphi_h(\varphi)$ dimensionless, Eqs. (A8) and (A10) yield

$$\psi_{h0} = \frac{c_{g0}}{c_{h0}} \frac{g_c}{l_c}, \quad \psi_{g0} = \frac{1}{c_{h0}c_{g0}} g_c l_c, \tag{A12}$$

where

$$c_{h0} = \text{sign}(\varphi_+ - \varphi_-) \int_{\varphi_-}^{\varphi_+} \frac{\psi_h(\varphi) + \Delta\psi_h(\varphi)}{\sqrt{\Delta\psi_h(\varphi)}} \, d\varphi, \quad c_{g0} = \frac{|\varphi_+ - \varphi_-|}{\sqrt{\Delta\psi_h(\varphi(0))}}. \tag{A13}$$

The homogeneous and gradient terms of the surface energy can be further expressed as

$$\psi_h = \frac{c_{g0}}{c_{h0}} \frac{g_c}{l_c} \psi_h(\varphi), \quad \psi_g = \frac{1}{c_{h0}c_{g0}} g_c l_c |\partial_x \varphi|^2. \tag{A14}$$

Using $\psi_h(\varphi(x)) = 1 - \varphi$ and substituting Eq. (A14) into Eq. (A1), Eq. (39) is obtained

$$\rho_0 \psi_f = \frac{g_c}{l_c}(1-\varphi) + \frac{1}{2} g_c l_c |\nabla\varphi|^2. \tag{A15}$$

**Appendix B. Mesh sensitivity validation**

Fig. B1 illustrates the influence of the filter coefficient $\eta$ on the numerical behavior of the strain gradient CP simulations. As shown in Fig. B1(a), when $\eta = 0$ (i.e., no filtering is applied), the stress-strain responses exhibit pronounced mesh sensitivity, with no sign of convergence as the mesh is refined. As $\eta$ gradually increases from 0 to 100, the overall strain rate decreases, as shown in Fig. B1(b). Figs. B1(c1)-(c4) display the spatial distributions of plastic slip for various $\eta$ values, while Figs. B1(d1)-(d4) show the corresponding GND density fields at a strain of 10%. In the absence of filtering ($\eta = 0$), severe mesh dependence is observed, manifested as highly localized slip bands and unrealistically concentrated GND clusters. With increasing $\eta$, both the plastic slip and GND density fields become progressively smoother, demonstrating the filter's effectiveness in suppressing non-physical oscillations. However, when the filtering strength becomes excessive (e.g., $\eta = 100$), the dislocation structures become overly diffused, leading to an underestimation of the mechanical response.

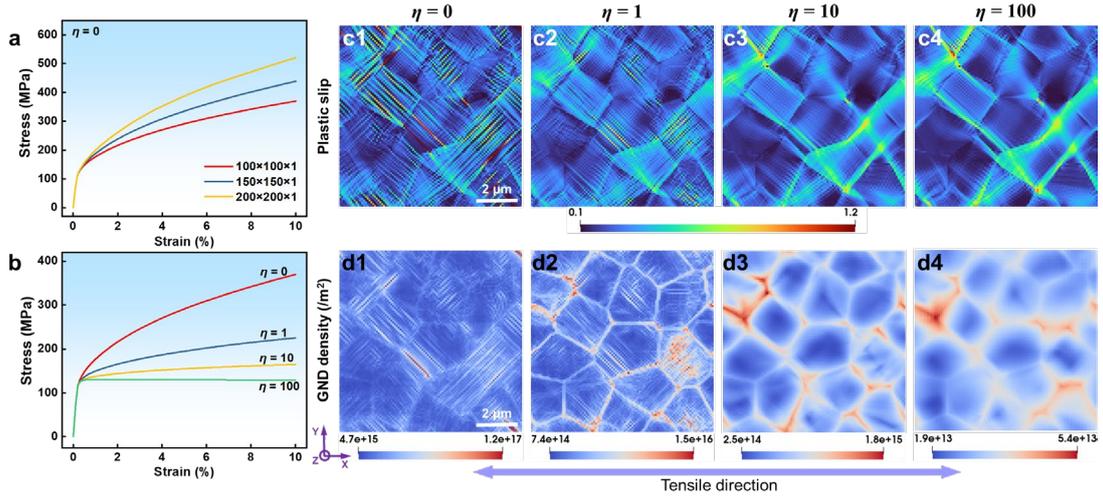

**Fig. B1.** (a) Stress-strain responses for different mesh resolutions (100×100×1, 150×150×1, and 200×200×1) when $\eta = 0$. (b) Stress-strain curves for different filter coefficients ($\eta = 0, 1, 10,$ and $100$). (c) Distributions of plastic slip at 10% strain for various $\eta$ values: (c1) $\eta = 0$, (c2) $\eta = 1$, (c3) $\eta = 10$, and (c4) $\eta = 100$. (d) Corresponding GND density distributions at 10% strain : (d1) $\eta = 0$, (d2) $\eta = 1$, (d3) $\eta = 10$, and (d4) $\eta = 100$. (average grain size $\approx$ 2 μm)

To further validate the numerical prediction of GND density, Fig. B2 compares the simulation result with the experimental characterization conducted at 10% uniaxial tensile strain for a microstructure with an average grain size of ~12 μm (Jiang et al., 2013). The simulated GND density field is obtained using a spectral filtering parameter of $\eta = 8$, yielding an average GND density of $2.03 \times 10^{14}$ m$^{-2}$, which is consistent with the experimentally observed value of ~$2.1 \times 10^{14}$ m$^{-2}$. Furthermore, the simulated GND distribution captures the key heterogeneities observed in the experiment, particularly the accumulation of GNDs near grain boundaries and lower densities in grain interiors. This agreement supports the physical fidelity of the adopted filtering scheme. Based on a balance between numerical stability and physical realism, a filter parameter of $\eta = 8$ is employed throughout the remaining simulations in this study.

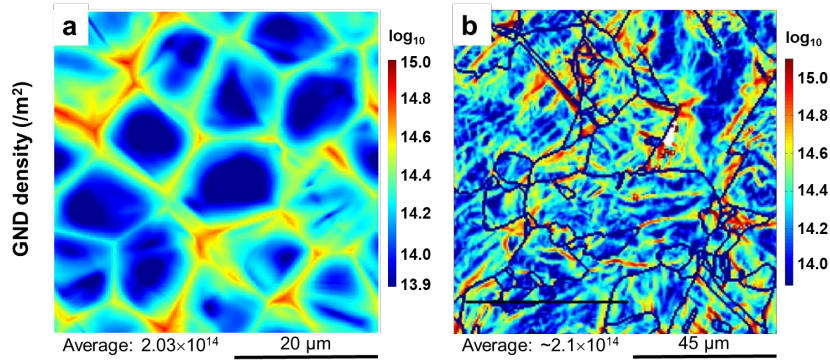

**Fig. B2.** (a) Simulated GND density map at 10% tensile strain using $\eta = 8$; (b) experimentally

characterized GND density map at 10% strain for a microstructure with an average grain size of ~12 μm (Jiang et al., 2013).

Fig. B3 investigates the influence of grain size and mesh density on the stress-strain response, considering diffusion regularization with a filter coefficient of $\eta = 8$. Two representative grain sizes (10 μm and 2 μm) are considered, each simulated under three mesh densities: 100×100×1, 150×150×1, and 200×200×1. As shown in Fig. B3(a), the larger grain size exhibits minimal mesh sensitivity, with all curves nearly overlapping. In contrast, Fig. B3(b) demonstrates that the smaller grain size leads to more pronounced mesh sensitivity due to increased gradient effects. Figs. B3(c) and (d) show the evolution of GND density for 10 μm and 2 μm grains, respectively. For the 10 μm case, the GND curves exhibit good agreement across all mesh densities, indicating reliable convergence. In the 2 μm case, the GND density shows slightly increased mesh dependence, especially at higher strains, due to stronger gradient contributions in smaller grains. Nevertheless, the variation remains within an acceptable range. Overall, smaller grains generate higher GND densities and exhibit enhanced strain hardening behavior, consistently captured across different mesh resolutions.

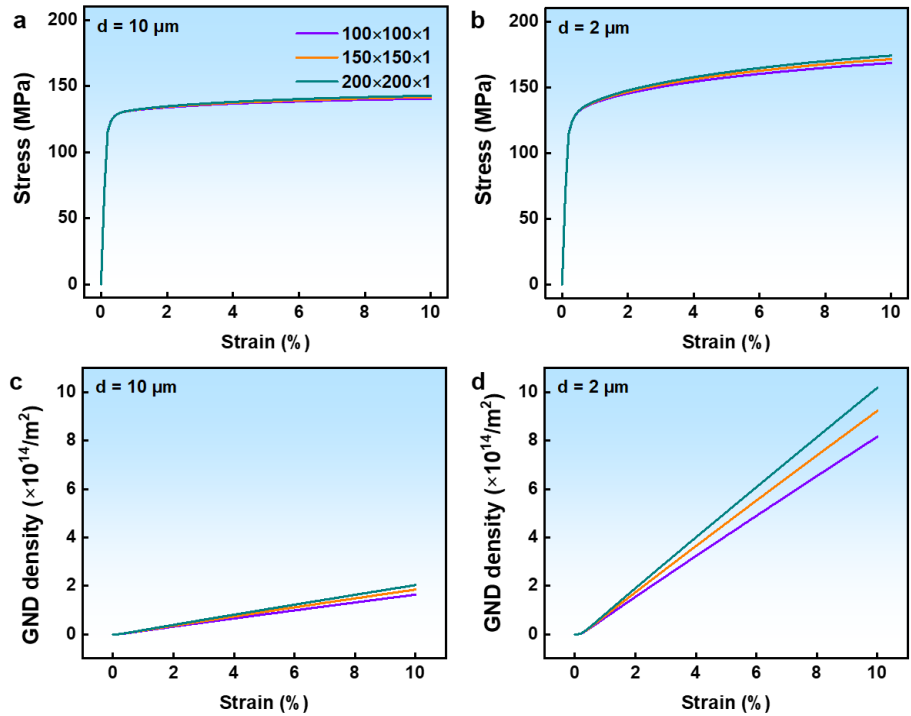

**Fig. B3.** Mechanical response and dislocation evolution with filter coefficient $\eta = 8$ for different grain sizes: (a) stress-strain curves for 10 μm grains, (b) stress-strain curves for 2 μm grains, (c) GND density evolution for 10 μm grains, and (d) GND density evolution for 2 μm grains. Three mesh resolutions are investigated: 100×100×1 (coarse), 150×150×1 (medium), and 200×200×1 (fine).

Fig. B4 presents the spatial distributions of plastic slip and GND density at 10% strain for the polycrystal with a grain size of 2 μm under three different mesh densities, with the filter coefficient fixed at $\eta = 8$. As shown in Figs. B4(a1)-(a3), the overall morphology and intensity of plastic slip remain highly consistent across all mesh densities, demonstrating the robustness of the plastic deformation prediction. The GND density distributions in Figs. B4(b1)-(b3) exhibit slightly increased resolution and magnitude with finer meshes; however, these variations remain within an acceptable range and do not compromise the physical interpretability of the results. These findings confirm that the applied regularization effectively mitigates excessive mesh sensitivity while maintaining the fidelity of the predicted plastic slip and GND evolution, thereby ensuring reliable characterization of strain gradient effects.

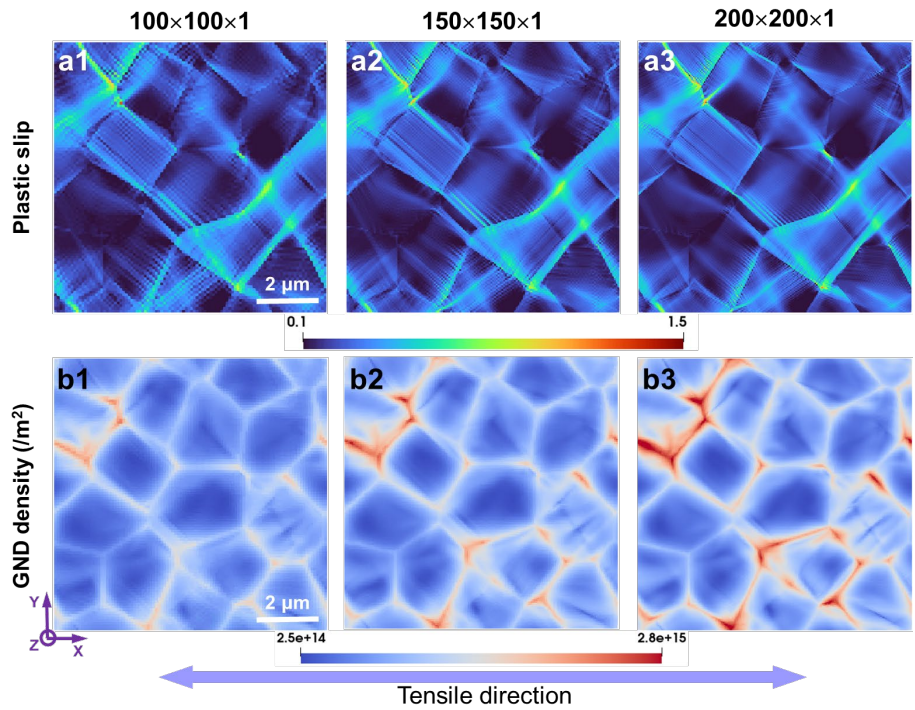

**Fig. B4.** Distribution of internal variables ($\eta = 8$) for a 2 μm grain size polycrystal at 10% strain under three mesh resolutions: (a1-a3) plastic slip and (b1-b3) GND density distribution, comparing mesh densities of 100×100×1 (coarse), 150×150×1 (medium), and 200×200×1 (fine).

**Appendix C. Numerical modelling of damage evolution: Mesh sensitivity and effects of periodic boundary conditions**

Damage model generally exhibits stronger mesh sensitivity compared to plasticity simulations, and this effect may become particularly pronounced in strongly anisotropic polycrystalline materials. We have therefore conducted a mesh sensitivity study of our damage model. Fig. C1 presents crack morphologies obtained at the same overall strain level (11.7%) for three different mesh densities: coarse (150×150×1), medium (200×200×1), and fine (250×250×1). The results show that the crack morphology obtained with the medium and fine meshes is almost identical, whereas the coarse mesh produces noticeable deviations in both crack length and crack location. Considering the balance between computational efficiency and numerical accuracy, the medium mesh (200×200×1) has been adopted for all subsequent simulations.

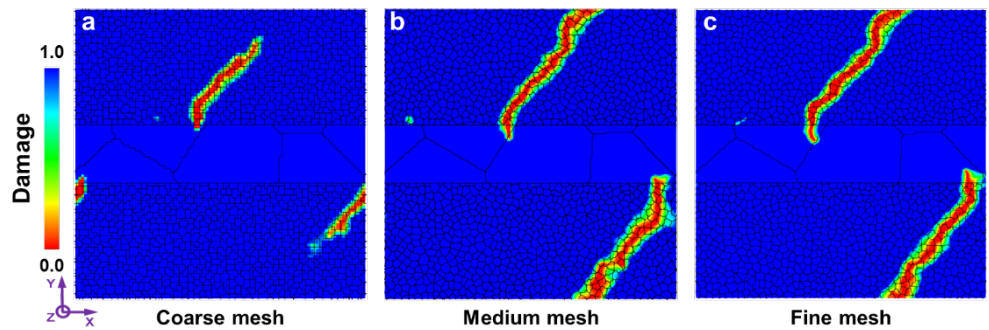

**Fig. C1.** Crack morphologies at a strain of 11.7% for different mesh densities: (a) coarse mesh (150×150×1), (b) medium mesh (200×200×1), and (c) fine mesh (250×250×1).

The application of PBCs inevitably influences the crack morphology, since one simulates not a single crack but rather a periodic array of mutually interacting cracks, hence, the number of microcracks is artificially enhanced by the periodic images and reflects the size of the periodic supercell rather than any physical process. To examine to which extent this affects our conclusions regarding crack morphology and propagation mechanisms, we have constructed a larger model—four times the original size (100×100×0.25 $\mu m^3$) as shown in Figs. C2(a1-a4). Evidently, the influence of the periodic BC on the nucleation and propagation of the emerging crack is reduced relative to the smaller model. However, the global fracture scenario is unchanged: Cracks nucleate at the HI (Fig. C2(a1)) and propagate through the FG region (Fig. C2(a2)). These cracks are arrested at the CG-FG interface, while further cracks nucleate and propagate in the adjacent FG region on the other side of the CG layer (Figs. C2 (a3,a4)). Ultimately, stress concentrations at the HI enable crack propagation through the CG region and cause system failure. We now use one quarter of the enlarged model which contains the crack nucleation site (dashed quadrant in Fig. C2(a1)), and continue this periodically. In this case, the nucleation and propagation process is very similar (Figs. C2(b1-b4)), with the evident difference that the smaller supercell, owing to the PBCs, leads to a total crack length per unit area that is twice as high. As a consequence, the smaller supercell requires a correspondingly higher work of fracture, which leads to a higher stress level in the softening region, as seen from comparison of the stress-strain curves in Fig. C2(c). We conclude that, while the stress level in the softening region of the stress-strain curve depends on the supercell size, our conclusions regarding the fracture mechanism are robust in this respect.

A second mechanism how PBCs affect the failure behavior resides in the elastic self-interaction of the nucleating crack. If we use a supercell of aspect ratio 1:1, this self interaction is enhanced since cracks tend to propagate under an angle of about 45 degrees, following the direction of preceding shear bands. Such cracks then close onto themselves across the PBCs, which accelerates their propagation as the crack-tip stress concentrations mutually reinforce.

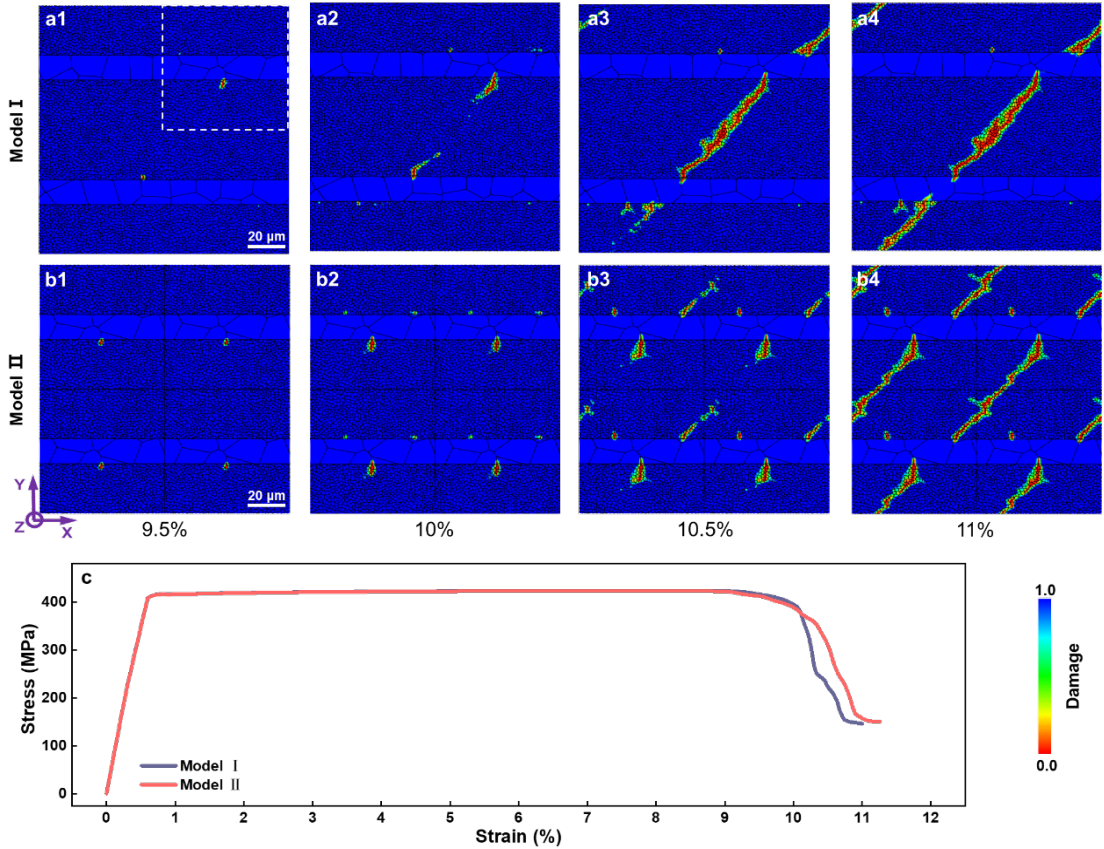

**Fig. C2.** (a1-a4) Strain evolution of the crack morphology in a larger-scale RVE (100×100×0.25 µm$^3$); (b1-b4) strain evolution of crack morphology in a periodically replicated smaller supercell (dashed square in (a1)) containing the crack nucleation site; (c) stress strain curves for both models.

This effect can be changed by modifying the aspect ratio of the periodic supercell to a decidedly irrational ratio of 1:1.618 as shown in Fig. C3. In this case, the initial damage evolution and stress-strain curve are nearly identical to that of a model with commensurate aspect ratio 1:1. However, as soon as the cracks start to self-interact across the PBCs, differences appear: The commensurate model experiences a strong stress drop as the stress concentrations of the crack and its periodic image mutually reinforce, leading to rapid crack growth (Fig. C3(a)). In the incommensurate model, on the other hand, the periodic image propagates into a zone where the stress is reduced by the presence of the initial crack, and therefore crack propagation and stress drop proceed in a much slower manner.

Our observations regarding the effect of PBCs can be summarized as follows: the qualitative scenario concerning crack nucleation and initial crack propagation in the heterostructure as described in the main paper is robust and does not strongly depend on the PBCs set-up. Cracks nucleate on the FG side of the HIs, propagate across the adjacent FG region with concomitant arrest at the CG region, and only in a second step penetrate the CG region. This scenario is robust upon increases of the supercell size and/or changes of the supercell aspect ratio. The stress-strain curve in the softening regime, on the other hand, changes when the supercell size is increased (more rapid stress decrease because the specific areal density of microcracks in the supercell and its periodic images is lower for a larger supercell) (Fig. C3(b)). It also changes when the supercell aspect ratio and thereby the interaction between the crack tip and its periodic image are changed (less rapid crack growth and stress decrease for an incommensurate supercell). We conclude that comparisons of the stress-strain behavior in the softening regime are only feasible between samples where the PBCs are set-up in an identical manner, and even then can be made only in a qualitative sense.

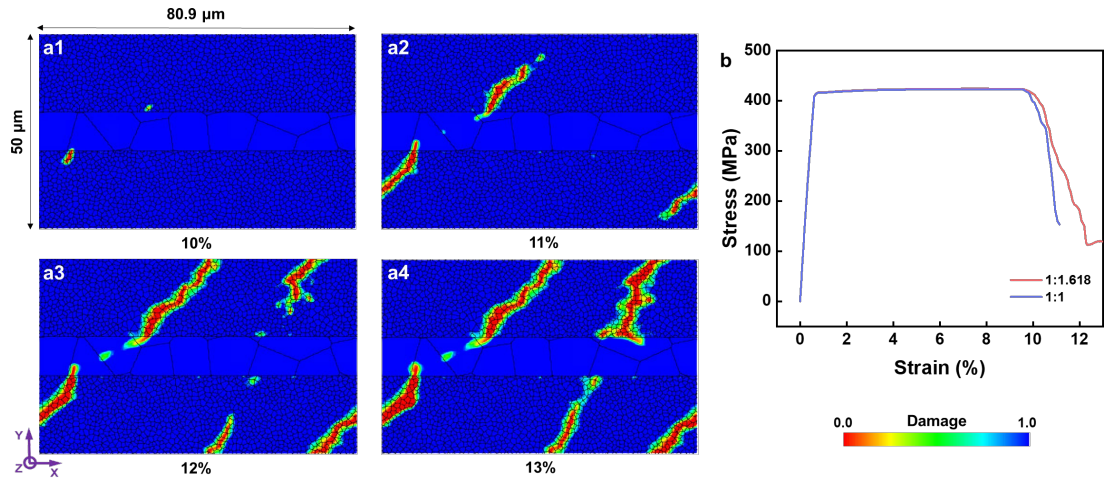

**Fig. C3.** (a1-a4) Strain evolution of the Crack morphology in a RVE with incommensurate aspect ratio (50×80.9×0.25 μm$^3$); (b) stress-strain curves for the original and the incommensurate model.